%% file: main.tex
\documentclass{article}

    \PassOptionsToPackage{numbers, compress}{natbib}

 \usepackage[preprint]{neurips_2025}

\usepackage[utf8]{inputenc} 
\usepackage[T1]{fontenc}    
\usepackage[pagebackref=true,breaklinks,colorlinks,citecolor=blue,linkcolor=blue,urlcolor=blue,hypertexnames=false]{hyperref}
\usepackage{url}            
\usepackage{booktabs}       
\usepackage{amsfonts}       
\usepackage{nicefrac}       
\usepackage{microtype}      
\usepackage{xcolor}         


\input{macro}

\usepackage{booktabs}
\usepackage{array}
\usepackage{multirow}
\usepackage{makecell}
\usepackage{wrapfig}
\usepackage{graphicx}
\usepackage{caption}
\usepackage{wrapfig}
\usepackage{amsmath}
\usepackage[font=footnotesize,skip=3pt,subrefformat=parens]{subcaption}

\title{DGS-LRM: Real-Time Deformable 3D Gaussian Reconstruction From Monocular Videos}

\author{
Chieh Hubert Lin$^{1,2}$ \hspace{.8em} 
Zhaoyang Lv$^1$ \hspace{.8em}
Songyin Wu$^{1,3}$ \hspace{.8em}
Zhen Xu$^1$ \hspace{.8em}
Thu Nguyen-Phuoc$^1$ \\
\textbf{
Hung-Yu Tseng$^1$ \hspace{1em}
Julian Straub$^1$ \hspace{.7em}
Numair Khan$^1$ \hspace{.7em}
Lei Xiao$^1$ \hspace{.7em}
Ming-Hsuan Yang$^{1,2}$} \\
\textbf{
Yuheng Ren$^1$ \hspace{.7em}
Richard Newcombe$^1$ \hspace{.7em}
Zhao Dong$^1$ \hspace{.7em}
Zhengqin Li$^1$} \\[.5em]
$^1$Meta \hspace{1em} 
$^2$UC Merced \hspace{1em} 
$^3$UC Santa Barbara
}

\begin{document}

\onecolumn{%
\maketitle
\vspace{-2.0em}
\renewcommand\twocolumn[1][]{#1}%
    \includegraphics[width=\linewidth]{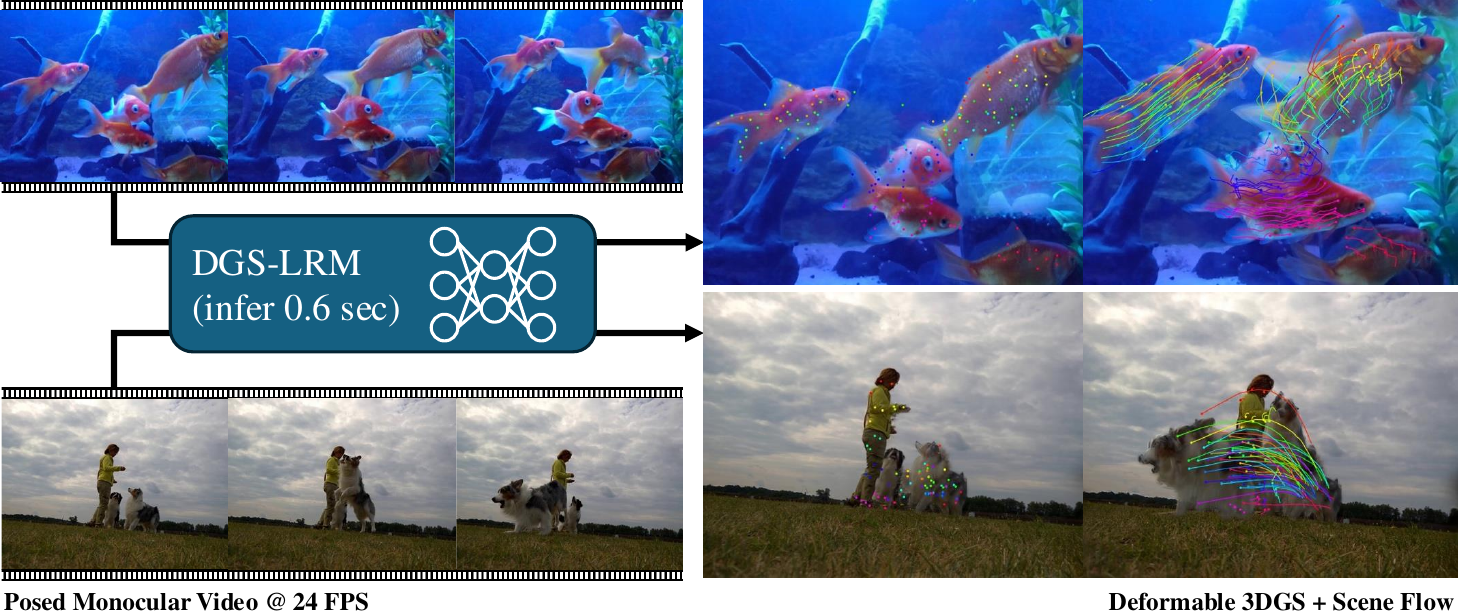}
    \vspace{-1em}
    \captionsetup{hypcap=false}\captionof{figure}{
        Our proposed Deformable Gaussian Splats Large Reconstruction Model (DGS-LRM) takes posed monocular videos as input and predicts deformable 3D Gaussians in a single feedforward pass. On the right, we render novel views from the predicted deformable Gaussians and sample 2D trajectories from the aggregated 3D scene flow of deformable 3D Gaussians.
    } 
    \label{fig:teaser}
    \vspace{1.2em}
}

\input{1-abstract}
\input{2-intro}

\input{3-related}
\input{4-method}

\input{5-exp}

\input{6-conclusion}
{
    \small
    \bibliographystyle{ieeenat_fullname}
    \bibliography{main}
}

\end{document}

%% file: macro.tex
\usepackage{color}
\definecolor{crimson}{rgb}{0.86, 0.08, 0.24}
\definecolor{gray}{rgb}{0.5,0.5,0.5}
\definecolor{green}{rgb}{0, 0.4, 0}
\definecolor{orange}{rgb}{1, 0.5, 0}
\definecolor{mahogany}{rgb}{0.75, 0.25, 0.0}
\definecolor{purple}{rgb}{0.6, 0, 0.6}
\definecolor{darkgreen}{rgb}{0, 0.4, 0}
\definecolor{frenchblue}{rgb}{0.0, 0.45, 0.73}
\definecolor{red}{rgb}{1,0,0}
\definecolor{yellow}{rgb}{1,1,0}
\definecolor{magenta}{rgb}{1,0,1}
\definecolor{pink}{rgb}{1,0.412,0.706}
\definecolor{newgreen}{rgb}{0, 0.6, 0.2}
\definecolor{lightred}{HTML}{fff3f3}

\makeatletter
\DeclareRobustCommand\onedot{\futurelet\@let@token\@onedot}
\def\@onedot{\ifx\@let@token.\else.\null\fi\xspace}

\makeatother

\newlength\paramargin
\newlength\figmargin
\newlength\subfigmargin
\newlength\presecmargin
\newlength\secmargin
\newlength\subsecmargin
\newlength\subsubsecmargin
\newlength\tabmargin
\newlength\eqmargin
\newlength\paraskip

\long\def\ignorethis#1{}

\newcommand{\Paragraph}[1]
{\vspace{1mm} \noindent\textbf{#1}}

\newcommand{\comment}[1]{}

%% file: 1-abstract.tex
\begin{abstract}

We introduce the {\underline D}eformable {\underline G}aussian \underline{S}plats {\underline L}arge {\underline R}econstruction {\underline M}odel (DGS-LRM), the first feed-forward method predicting deformable 3D Gaussian splats from a monocular posed video of any dynamic scene. Feed-forward scene reconstruction has gained significant attention for its ability to rapidly create digital replicas of real-world environments. However, most existing models are limited to static scenes and fail to reconstruct the motion of moving objects. Developing a feed-forward model for dynamic scene reconstruction poses significant challenges, including the scarcity of training data and the need for appropriate 3D representations and training paradigms. To address these challenges, we introduce several key technical contributions: an enhanced large-scale synthetic dataset with ground-truth multi-view videos and dense 3D scene flow supervision; a per-pixel deformable 3D Gaussian representation that is easy to learn, supports high-quality dynamic view synthesis, and enables long-range 3D tracking; and a large transformer network that achieves real-time, generalizable dynamic scene reconstruction. Extensive qualitative and quantitative experiments demonstrate that DGS-LRM achieves dynamic scene reconstruction quality comparable to optimization-based methods, while significantly outperforming the state-of-the-art predictive dynamic reconstruction method on real-world examples. Its predicted physically grounded 3D deformation is accurate and can be readily adapted for long-range 3D tracking tasks, achieving performance on par with state-of-the-art monocular video 3D tracking methods.

\end{abstract}

%% file: 2-intro.tex
\vspace{\secmargin}
\section{Introduction}
\vspace{\secmargin}

Reconstructing a dynamic scene from a monocular video, recovering accurate geometry, appearance, and motion, remains a significant challenge in computer vision and graphics. This task has numerous applications, including visualization, augmented/virtual reality (AR/VR), and robotics. Recent advances in this domain \cite{li2023dynibar,wang2024shape,yang2024deformable} have been largely driven by the development of neural representations, such as neural radiance fields \cite{mildenhall2021nerf} and 3D Gaussian splats \cite{kerbl20233d}, as well as deep priors for specific scene attributes like depth \cite{wang2024dust3r,leroy2024grounding,li2024megasam,yang2024depth,ranftl2021vision} and flow \cite{teed2020raft,teed2021raft,karaev2024cotracker}. These methods tackle dynamic scene reconstruction by optimizing particular scene representations using densely captured images, integrating various deep priors to provide robust regularizations. Although recent methods using variants of Gaussian splatting \cite{kerbl20233d} representation can achieve real-time rendering speeds \cite{yang2024deformable,wu20244d,duan20244d,wang2024shape}, the optimization process is often time-consuming and computationally expensive, limiting their practical applicability.

The recent development of generalizable 3D feed-forward networks \cite{wang2024dust3r,zhang2024monst3r,hong2024lrm,zhang2025gs,xu2024grm} directly predicts 3D representations from sparse-view image inputs, achieving speeds several orders of magnitude faster than previous optimization-based methods. Inspired by the success of foundation models in natural language processing \cite{radford2018improving} and 2D computer vision \cite{dosovitskiy2020image,kirillov2023segment}, recent works explore training transformer-based \cite{vaswani2017attention} networks on large-scale 3D \cite{deitke2023objaverse,deitke2024objaverse} or video datasets to enable generalizable 3D reconstruction. However, prior efforts assume static scenes, leaving the challenge of handling dynamic scenes and accurately predicting motion still unsolved.

In this paper, we present DGS-LRM, the first feed-forward transformer designed to predict deformable 3D Gaussians from a posed monocular video. The predicted deformable 3D Gaussians can render accurate geometry, appearance, and scene flow of dynamic scenes. We introduce three key technical innovations to address this challenge, including a deformable 3D representation, an enhanced synthetic dataset that can train this method at scale, and an end-to-end network architecture. 

First, we employ per-pixel deformable 3D Gaussians to represent predicted dynamic scenes, extending the recent success in feed-forward static 3D Gaussians prediction methods \cite{zhang2025gs,xu2024grm,charatan2024pixelsplat}. For each pixel in a video frame, our method predicts its corresponding 3D Gaussian splat and its 3D scene flow from the current timestamp to all other timestamps in the input video, enabling high-quality novel view synthesis through warping 3D Gaussians across frames. Additionally, our representation is robust to occlusions and discontinuities. The predicted 3D scene flow can be chained together using sliding windows, achieving performance on par with the state-of-the-art 3D tracking method \cite{xiao2024spatialtracker}.


Second, motivated by the recent trend in training LRM with pure synthetic data~\cite{xie2024lrm,jiang2024megasynth}, we utilize multi-view synthetic data with ground-truth 3D scene flow as a primary supervision during training. This contrasts with prior scene-level feed-forward reconstruction models, which are typically trained on monocular video or RGB-D images. 
However, using pure photometric supervision on a monocular video has ambiguities in motion and geometry predictions for dynamic scenes.
We create a customized large-scale dataset using Kubric \cite{kubric}, featuring multi-view renderings paired with per-pixel 3D scene flow. 
Our results show that training on such a large-scale multi-view synthetic dataset significantly enhances reconstruction quality and enables good generalization on real-world data as well. 

Third, we employ temporal tokenization \cite{sora,polyak2025moviegencastmedia} to compress input videos into compact, small pixel cubes. Unlike GS-LRM \cite{zhang2025gs}, which tokenizes each frame independently, our approach is computationally efficient and scales well for both training and inference. 
Additionally, we incorporate discretely sampled temporally distant reference frames as inputs to leverage frames with larger camera baselines, effectively reducing geometric ambiguities.

Our best model achieves real-time inference on an A100 GPU while maintaining reconstruction quality comparable to optimization-based deformable Gaussian splatting methods \cite{yang2024deformable}, which require hours of computation and several pre-trained models for initialization. Compared to the existing predictive dynamic object reconstruction method \cite{ren2024l4gm}, DGS-LRM demonstrates significantly better generalization ability on real-world examples, improving PSNR by 3 points on \cite{dycheck} when we segment out dynamic objects in both inputs and outputs for a fair evaluation. Moreover, DGS-LRM accurately reconstructs object motion. Our predicted 3D scene flow delivers competitive quantitative performance \cite{xiao2024spatialtracker} on standard point tracking benchmarks \cite{kubric,poointodyssey}, indicating that the generated deformation is continuous and 3D accurate. More qualitative results from DAVIS~\cite{perazzi2016benchmark} further demonstrate high-quality novel-view synthesis and accurate flow prediction, as shown in Figure \ref{fig:teaser}.

%% file: 3-related.tex
\vspace{\secmargin}
\section{Related Work}
\vspace{\secmargin}

\Paragraph{Dynamic reconstruction and view synthesis}
Early dynamic scene reconstruction methods \cite{bozic2020deepdeform,innmann2016volumedeform,dou2016fusion4d,newcombe2015dynamicfusion,zollhofer2014real} primarily focus on non-rigid mesh reconstruction, typically requiring RGBD images as inputs. Video depth prediction methods predict consistent depth maps across video frames \cite{kopf2021robust,luo2020consistent,zhang2022structure,zhang2021consistent} by integrating monocular depth priors with strong hand-crafted regularizations. Both of these approaches concentrate on geometry and are not equipped to support realistic novel view synthesis. Recent advancements in dynamic view synthesis have shown significant progress by incorporating novel neural representations, such as neural radiance fields \cite{mildenhall2021nerf} and 3D Gaussian splats \cite{kerbl20233d}, into dynamic scene reconstruction. Most of these methods require multi-view videos as inputs \cite{lombardi2019neural,broxton2020immersive,cao2023hexplane,fridovich2023k,li2022neural,li2024spacetime,song2023nerfplayer,stich2008view,wang2022fourier}, which considerably limits their practical applicability. Several methods \cite{liu2023robust,du2021neural,duan20244d,li2021neural,park2021nerfies,park2021hypernerf,wang2021neural,wu20244d,xian2021space,yang2024deformable,yang2023real,wang2024shape} target the more challenging task of monocular dynamic view synthesis. While these methods remarkably improve view synthesis quality, they heavily rely on geometry and motion priors from pre-trained models, alongside meticulously designed time-consuming optimization processes to achieve state-of-the-art reconstruction quality. On the contrary, DGS-LRM aims to learn priors from a single large transformer model, enabling efficient, generalizable dynamic scene reconstruction in a feed-forward manner. 

\Paragraph{Feed-forward reconstruction} Many feed-forward methods can predict neural 3D representations for novel view synthesis or geometry reconstruction. 
Early methods \cite{chen2021mvsnerf,yu2021pixelnerf,wang2021ibrnet,xu2022point} often struggle to match the reconstruction quality of optimization-based methods and, therefore, require fine-tuning \cite{xu2022point,chen2021mvsnerf} to enhance their results. Recently, many works \cite{wang2024dust3r,leroy2024grounding,chen2024mvsplat} have significantly increased network capacity by employing large transformer networks for better 3D reconstruction. Among these, large reconstruction models (LRMs) \cite{hong2024lrm,li2023instant3d,xu2023dmv3d,wei2024meshlrm,wang2023pf,zhang2025gs} represent a family of approaches that achieve state-of-the-art novel view synthesis quality from sparse, posed, or unposed images. The latest LRM methods \cite{zhang2025gs,xu2024grm} predict pixel-aligned 3D Gaussians for realistic static scene-level view synthesis but fail to handle dynamic objects. Several concurrent works aim to reconstruct dynamic scene geometry \cite{zhang2024monst3r,li2024megasam, wang2025cut3r} but not appearance and motion. The closest existing method \cite{ren2024l4gm} predicts time-dependent 3D Gaussians to reconstruct dynamic objects. However, we observe that it generalizes significantly worse on real-world examples, even when the dynamic objects are segmented out in the inputs. This may be due to the limited motion and appearance diversity in its training data and its 3D representation, which fails to model accurate scene flow. A concurrent work \cite{liang2024feed} achieves better generalization by training on self-curated internet videos, but it still does not reconstruct accurate scene flow as our model.



\Paragraph{Flow and tracking} The majority of tracking methods aim to find correspondences in 2D image space. Classical optical flow methods \cite{black1993framework,brox2009large,brox2004high,ilg2017flownet,dosovitskiy2015flownet,teed2020raft} estimate dense 2D pixel motion between two consecutive frames and, therefore, are not suitable for long-range tracking. Feature tracking methods \cite{bay2006surf,detone2018superpoint,lowe2004distinctive,rublee2011orb} can track pixels over long ranges, but only handle sparse points. Several efforts \cite{sand2008particle,rubinstein2012towards} have been made to combine the merits of the two to achieve dense long-range 2D tracking, either by concatenating consecutive 2D flows through test-time optimization \cite{neoral2024mft,wang2023tracking} or by relying on data-driven approaches \cite{doersch2023tapir,harley2022particle,karaev2024cotracker}. Another line of research estimates correspondences in 3D space to mitigate issues caused by 3D-2D projection and to leverage better regularization. Most of these approaches require RGBD images or point clouds as inputs \cite{gu2019hplflownet,liu2019flownet3d,puy2020flot,wang2020flownet3d++,jaimez2015primal,quiroga2014dense,sun2015layered,teed2021raft} or test-time optimization \cite{hur2020self,li2023dynibar,li2021neural} to jointly reconstruct geometry. SpatialTracker \cite{xiao2024spatialtracker} is the closest state-of-the-art work that predicts dense 3D scene flow from a monocular video. Experiments show that DGS-LRM achieves comparable tracking accuracy while offering a more versatile framework that supports high-quality novel view synthesis.

%% file: 4-method.tex
\vspace{\secmargin}
\section{Methodology}
\vspace{\secmargin}

\input{figures/pipeline}

In Sec.~\ref{sec:arch}, we introduce the network architecture and the convention of DGS-LRM.
Sec.~\ref{sec:loss} outlines key training details.
Sec.~\ref{sec:data} describes synthetic training data.

\vspace{\subsecmargin}
\subsection{DGS-LRM}
\vspace{\subsecmargin}
\label{sec:arch}
Figure \ref{fig:pipeline} illustrates the inputs, transformer network architecture, and the predicted 3D representation of DGS-LRM, which we will discuss in detail. 

\Paragraph{Network inputs.}
Given a monocular video sequence $\mathcal{I}=\{I_0, \dots, I_N\}$ with per-frame camera parameters $\mathcal{C}=\{C_0, \dots, C_N\}$, DGS-LRM aims to predict a deformable Gaussian splatting (DGS) reconstruction of the visible environment.
Following LRM~\cite{hong2024lrm}, we encode camera calibrations as Plücker rays, where each $C_n \in \mathcal{C}$ pairs with an image  $I_i \in \mathcal{I}$ with the same resolution ($H \times W$).
Each pixel $p$ of $C_n$ is a seven-dimensional vector containing a Plücker ray and a timestamp.
%
%
The timestamp is the temporal index of the frame within the input time window $\{0, \dots, N\}$, and normalized to $\left[0, 1\right]$.
%

Processing a large number of video frames using a standardized self-attention transformer requires prohibitively large GPU VRAM. 
However, limiting the window size $N$ can reduce the camera baselines among input frames, making it more challenging for the model to reconstruct accurate geometry. 
%
%
To address this issue, we additionally introduce \textit{optional} reference frames $\mathcal{R}=\{R_0, \dots, R_K\}$ (omitting the corresponding Plücker rays for clarity).
These reference frames aim to sample more views with larger camera baselines, which provides additional clues for geometry reconstruction. 
Consequently, we sample $\mathcal{R}$ temporally distant from $\mathcal{I}$, and we do not predict per-pixel deformable 3D Gaussians for these frames.

\begin{wrapfigure}[14]{r}{0.5\textwidth}
    \centering
    \vspace{-1em}
    \includegraphics[width=\linewidth]{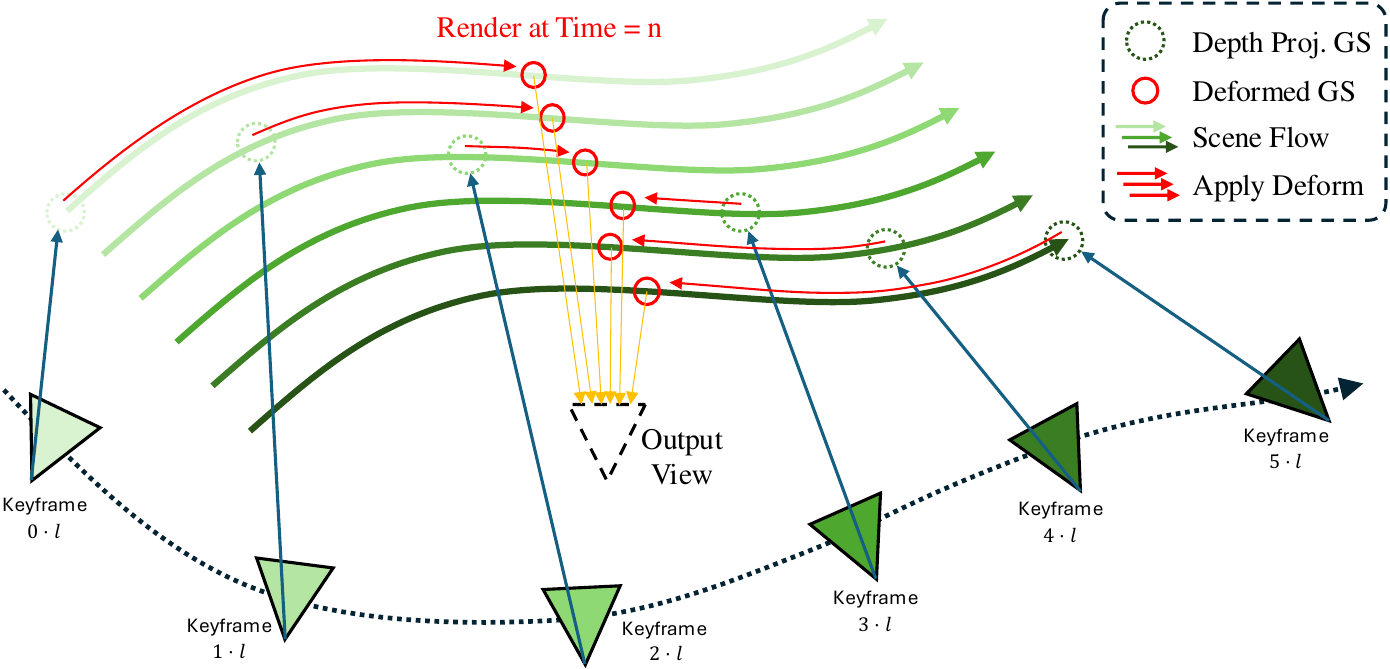}
    \caption{A visualization of pixel-aligned deformable 3D Gaussians are used for novel view synthesis at arbitrary timestamp $n$.}
    \label{fig:flow}
    \vspace{-0.1in}
\end{wrapfigure}
%
\Paragraph{Deformable Gaussian Splats.}
%
Given posed multiview images, DGS-LRM predicts a set of deformable Gaussian parameters by 
%
\begin{equation}
    \mathcal{G} = \text{DGS-LRM}(\mathcal{I}, \mathcal{C}, \mathcal{R}) \,\, , 
\end{equation}
where $\mathcal{G}=\{G_0, G_l, G_{2\cdot l}, \dots, G_N\}$. Here, $l$ is the temporal downsampling rate, and $G_{x\cdot l}$ is a set of deformable 3D Gaussians that are pixel-aligned with a keyframe $x\cdot l \in \{0,\dots,N\}$.
%
%
In $G_{x\cdot l}$, each pixel $p$ contains a single Gaussian splat $g_p$, parametrized by 1-channel depth $d_p$, 3-channel RGB colors, 4-channel quaternion rotation, 3-channel scale, 1-channel opacity, and a set of 3-channel deformation vectors $\mathbf{f}_p=\{f_0, f_1,\dots, f_N\}$ to warp Gaussian splat $g_p$ into the target timestamp $n\in\{0,...,N\}$. Our deformation vectors only model translation, and we observe that deforming other attributes such as rotation or opacity will not further improve reconstruction quality. To render a novel view image at timestamp $n$, we warp every Gaussian splats $g_p$ in $\mathcal{G}$ using the corresponding deformation vector $\mathbf{f}_p$ to form a new set of 3D Gaussian $\mathcal{W}_n$ for rendering
\begin{equation} 
\small
\mathcal{W}_n = \sum^{\mathcal{G}} \sum_{p} \text{warp}(g_p, \mathbf{f}_p) \,\, .
\end{equation}
A visualization of the warping process is shown in Figure \ref{fig:flow}. 
%
%

\Paragraph{Architecture.}
Our model consists of three components: an input tokenizer, a transformer, and output MLPs to project the results.
The input images (both $\mathcal{I}$ and $\mathcal{R}$) and their corresponding Plücker rays are first channel-concatenated together.
%
%
We use \textit{temporal tokenization}, inspired by
%
the temporal autoencoder from MovieGen~\cite{polyak2025moviegencastmedia}.
Instead of patchifying each image patch with $s \times s$ pixels, we consider a video as a volume and patchify a cube across spatial and temporal domains with $s \times s \times l$ pixels. This cube is turned into a token through a simple linear layer. 
%
We use $l=4$ in practice, reducing the number of tokens by four times, 
which significantly accelerates both training and inference.
In practice, the model is not trainable and expensive to infer without temporal tokenization.
We use a standard transformer architecture~\cite{vaswani2017attention} with 24 multi-head self-attention blocks.
%
%
We additionally add weight normalization~\cite{weight_normalization} to all parameters and observe that it can stabilize the training. Finally, we use two-layer MLPs to project the tokens into deformable GS parameters.

%
%
%

%
%
%


\vspace{\subsecmargin}
\subsection{Training}
\vspace{\subsecmargin}
\label{sec:loss}

%
\Paragraph{Photometric losses}
Similar to prior works \cite{zhang2025gs,xu2024grm}, DGS-LRM can be trained end-to-end with photometric losses through differentiable rasterization.
We sample a set of ground-truth images $\bar{\mathcal{I}} = \{\bar{I}_0, \dots, \bar{I}_Q\}$ with corresponding camera poses $\mathcal{\bar{C}}={\bar{c}_1, \dots, \bar{c}_Q}$, and define $n_q \in \{1, \dots, N\}$ as their timestamps.  
%
To produce a predicted rendering $\hat{I}_q$, we first deform the predicted 3D Gaussians to timestamp $n_q$ as $\mathcal{W}_{n_q}$, then rasterize it to the output camera pose $\bar{c}_q$ with
%
\begin{equation} 
\small
    \hat{I}_q = \text{rasterize}(\mathcal{W}_{n_q}, \bar{c}_q) \,\, .
\end{equation}
We then use MSE loss and perceptual loss \cite{lpips} to supervise the rendered image. 
\begin{equation}
    L_{\text{mse}} = \sum_{q}\text{MSE}(\bar{I}_q, \hat{I}_q)
    \,
    ; 
    \,\, 
    L_{\text{lpips}} = \sum_{q}\text{LPIPS}(\bar{I}_q, \hat{I}_q) \,\, .
\end{equation}

\vspace{-1em}
\Paragraph{Output view selection and dual-view supervision.}
%
Static scene LRMs~\cite{hong2024lrm, zhang2025gs} sample intermediate video frames from monocular videos as training data, which can lead to geometry and motion ambiguity when training with dynamic monocular videos.
%
%
%
%
%
%
%
%
To remove this ambiguity, we use time-synchronized multi-view videos to train DGS-LRM.

We start with sampling two video sequences; one serves as the input $\mathcal{I}$, while we sample $K$ frames from the other sequence as $\mathcal{R}$.
For output view supervision, we empirically found that sampling two views at the same timestamp provides significantly better training.  convergence
We name this strategy dual-view sampling and show its effectiveness in Figure~\ref{fig:dual-view-compare}.
With dual-view sampling, we sample two output video sequences instead of one and sample $Q/2$ frames from each stream at the shared timestamps.
To ensure the inputs and outputs of GDS-LRM have sufficient covisibility, we allow one of the output sequences to overlap with either the input or the reference sequence.


\Paragraph{Depth and scene flow supervisions.}
%
We apply depth supervision $L_\text{depth}$ to improve surface geometry
\begin{equation} 
\small
    L_\text{depth} = \sum^{\mathcal{G}}\sum_{p}\| \bar{d}_p - d_p \|_1 .
\end{equation}
Here, we directly supervise with pixel-aligned ground-truth depth values $\bar{d}_p$ for key frame $\{0, l, 2l,\dots,N\}$, which we observe to be more effective compared to supervising rendered depth maps at output views. Simply combining rendering loss and depth loss is still not enough for accurate motion reconstruction. We often observe discontinuities in the predicted deformation trajectories. Sometimes, the 3D Gaussian splats are even moved outside the camera frustum.  Consequently, we introduce the flow loss $L_{\text{flow}}$ to regularize the point deformation. Similar to the depth loss, for every key frame, we supervise our deformation vector $\mathbf{f}_p$ with ground-truth deformation to every timestamp. 
The flow loss is written as
\begin{equation} 
    \small
    L_{\text{flow}} = \sum^{\mathcal{G}}\sum_{p}\| \bar{\mathbf{f}}_p - \mathbf{f}_p \|_1 \,\, .
\end{equation}
How to obtain ground-truth deformation vectors $\bar{\mathbf{f}}_p $ will be detailed in Sec. \ref{sec:data}. 

%
%
%
%
%
%

\Paragraph{Total loss.}
Finally, the whole network is end-to-end trained with the total objective
\begin{equation}
    L_\text{total} = 
    L_\text{mse} +
    \lambda_\text{lpips} \cdot L_\text{lpips} +
    \lambda_\text{depth} \cdot L_\text{depth} +
    \lambda_\text{flow} \cdot L_\text{flow} \,\, ,
\end{equation}
where the $\lambda$'s are weighting factors of the objectives.
We use $\lambda_\text{lpips}=0.5$, $\lambda_\text{depth}=10$, and $\lambda_\text{flow}=10$.

\Paragraph{Scene normalization.}
We found that inconsistent scene scales in training and inference can cause instability and generalization issues.
%
%
%
We use a normalization approach similar to MegaSaM~\cite{li2024megasam} for both training and inference. We first use a monocular metric depth estimator \cite{piccinelli2024unidepth} to identify the scale of the scene and then normalize the scene scale so that the disparity of the 20th depth percentile is equal to 2.
Such a scene scale is applied to camera poses, ground-truth depths, and scene flows.



\vspace{\subsecmargin}
\subsection{Training Data Creation}
\label{sec:data}
\vspace{\subsecmargin}

Due to limited real-world \textit{posed} monocular videos that contain diverse \textit{dynamics} with sufficient ground truths (3D scene flow, multi-view images) to address the motion ambiguity, we primarily train our DGS-LRM on a self-generated Kubric~\cite{kubric} dataset. We also explored combining it with various real-world videos for training but did not observe major improvements.
%
%
%
%

\Paragraph{Customized Kubric.}
%
We follow the MOVi-E setting, using the Kubric engine~\cite{kubric} to create synthetic dynamic scenes. These scenes contain diverse objects being tossed around, simulated by a physics engine.
To ensure that our video is closer to real-world monocular dynamic videos \cite{dycheck}, we decrease the default maximum camera trajectory length from 8 meters to 0.5 meters to improve the model's generalization ability. 
%
We generate the dataset with 4 synchronized cameras.
We first sample one camera, then sample 3 other cameras relative to the first one, with distances ranging from 4 to 16 meters.
All these cameras share the same look-at points so that they can have sufficient co-visibility.
To further reduce the sim-to-real domain gap, we apply additional domain randomizations.
For each scene, we add motion blur to one of the cameras and make sure the camera is never being sampled as the output supervision but only as the input.
We also sample varying focal lengths for each camera, ranging from 25mm to 55mm.

\Paragraph{Scene Flow Extraction.}
We extract ground-truth 3D scene flow following MegaSaM~\cite{li2024megasam}.
The Kubric engine supports rendering per-pixel object coordinates, which specify the location where the ray cast from the camera intersects the object surface.
Moreover, Kubric also records the per-frame object trajectory and rotations during physics simulation.
Given that Kubric only contains rigid objects, we can retrieve the 3D scene flow for every pixel at every timestamp by combining the 2 attributes together.
%

However, 3D scene flow has an $O(N\times M\times P\times 3)$ space complexity.
In practice, using $N=24$, $M=6$, and $P=512\times512$ results in $0.84$ GB data for one camera.
Loading a batch of such data in each iteration would exhaust the memory and data I/O. 
Fortunately, we observe that the scene flow is significantly sparse, where the majority of points in the scene are stationary. By storing 3D scene flow as sparse tensors, we reduce the memory and I/O cost by 80\%. 
%
%
%
%
%
%

\vspace{\subsecmargin}
\subsection{Flow Chaining}
\label{method:chaining}
\vspace{\subsecmargin}

%
%
%

Our DGS-LRM is trained to handle short video clips (1s). Theoretically, it can be trained to handle longer videos, but this is computationally expensive in practice. However, we can \textit{chain} multiple sequences of scene flows together to achieve long-range tracking, which we will detail below.

Given two sequences with predicted scene flows, the \textbf{\textit{flow chaining}} first set the end frame of the first video as the first frame of the second video.
We deform the two independent deformable GS into the same timestamp.
Then, we find a nearest neighbor for each scene flow to temporally chain two flows into one.
The nearest neighbor is measured by two distance quantities: the distance between the deformed GS, and the direction similarity between the momentary scene flow.
For the first video, the momentary scene flow is the relative scene flow between the last two timestamps; for the second video, it is the relative scene flow between the first two timestamps.
The two distance quantities result in a six-value vector, and we measure the distance between these vectors with a simple L1 distance.
We use Faiss~\cite{faiss} to efficiently compute such distance with GPU support.
After computing all pair-wise distances between the scene flows from the first video and the scene flows from the second video, we index the nearest neighbor and temporally chain each pair of flows into one.

However, as some points in one video may be completely invisible in the second video (such as moving outside the frustum), some flows in one video may not have a valid match in the other video.
We use a threshold to filter out such cases and concatenate these unmatched scene flows with zero values, meaning the points remain stationary after it loses track.
These lost tracked scene flows are still evaluated when we compare them with the 3D tracking methods in Table~\ref{tab:quant-track-pod}.
We also provide an idealized evaluation where these unmatched scene flows are excluded from the evaluation, marked as Flow Valid (FV) in Table~\ref{tab:quant-track-pod}.

%% file: figures/pipeline.tex
\begin{figure*}
    \centering
    \includegraphics[width=\linewidth]{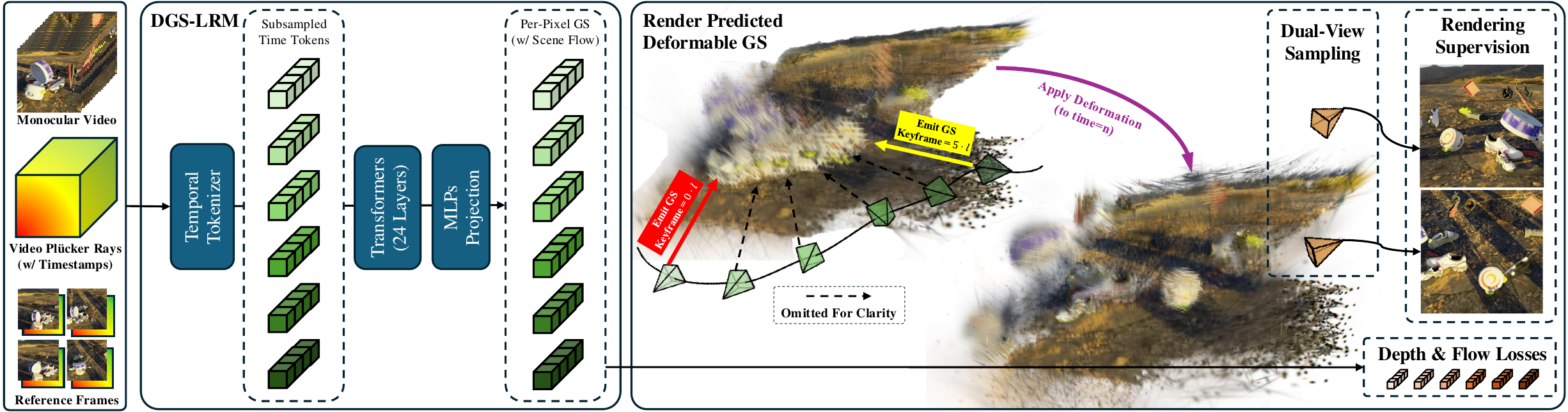}
    \vspace{-1.5em}
    \caption{
    \textbf{DGS-LRM overview.}
    We first concatenate multi-view videos with a time-aware Plucker ray and tokenize them using the spatial-temporal tokenizer. Then, the transformers take the sampled time tokens as input and predict per-pixel deformable Gaussians with 3D scene flow. During training, we rendered multi-view synthetic videos using Kubric. 
    We draw a dual-view ground-truth in each sample at the same timestamp, rendering views, depth, and scene flows.
    }
    \label{fig:pipeline}
    \vspace{-0.1in}
\end{figure*}

%% file: 5-exp.tex
\vspace{\secmargin}
\section{Experiments}
\vspace{\secmargin}

\input{tables/quant}

\input{figures/dycheck-nvs}
\input{figures/davis}

\Paragraph{Hyperparameters.}
We train our method with 64 H100 GPUs with 80GB VRAM.
For all variants of DGS-LRM, we use $N=24$ input frames with temporal sampling rate $l=4$, which results in 6 keyframes after temporal tokenization.
We use $K=4$ for reference views and set the number of output views per scene to $Q=8$.
For training efficiency, we first train the model at 256$\times$256 resolution and then fine-tune it at 512$\times$512 resolution.
We render the Kubric dataset according to these two setups and create 40{,}000 scenes (each with 4 synchronized cameras) for both resolutions.
%
%
For the first stage of training, we use a batch size of 15 per GPU, train for 40k iterations with a learning rate of $4e-4$, and then decay to $1e-6$ with a cosine learning rate scheduler.
For the second stage, we use a batch size of 8 per GPU, train for 20k iterations with a learning rate of $1e-4$, and then decay to $1e-6$ with a cosine learning rate scheduler.
For both stages, we use a learning rate warm-up for 500 iterations, which linearly ramps up the learning rate from 0 to the initial learning rate.
Similar to GS-LRM, we apply the common practice to save GPU VRAMs using xFormers~\cite{xFormers2022}, deferred backpropagation~\cite{deferred_backprop}, gradient checkpointing~\cite{gradient_checkpointing}, and BF16 mixed-precision training~\cite{bf_float16}.

\vspace{\subsecmargin}
\subsection{Novel Dynamic View Synthesis}
\label{exp:nvs}
\vspace{\subsecmargin}
We evaluate DGS-LRM on DyCheck~\cite{dycheck} and DAVIS~\cite{davis}. 
DGS-LRM requires the input to be a temporally continuous video with a non-stationary camera, and cannot process teleporting cameras with discrete poses.
Due to the lack of real-world benchmarks with multiple non-stationary and synchronized cameras, our quantitative evaluation is limited to DyCheck.
We use the iPhone subset of DyCheck, which includes two synchronized novel-view cameras for reconstruction metrics evaluation.
The iPhone subset contains 7 long monocular videos, each with 200-400 frames.
In addition, DyCheck also labels the covisibility between training and novel-view cameras and evaluates the masked version of reconstruction metrics.
DAVIS provides a large and diverse set of in-the-wild monocular videos to test the method's generalization.
We use MegaSaM~\cite{li2024megasam} to create the camera poses in DAVIS evaluation.
We only evaluate qualitative results, as DAVIS does not have novel-view cameras.

In Table~\ref{tab:quant-nvs-dycheck} and Figure~\ref{fig:nvs-dycheck}, we show that our DGS-LRM outperforms the baseline predictive method, L4GM \cite{ren2024l4gm}, while performing comparably to the state-of-the-art optimization-based reconstruction methods, which all take hours to complete.
L4GM is designed for 4D object reconstruction without reconstructing explicit deformation.
As L4GM only reconstructs dynamic foreground objects, we label the foreground mask following \cite{wang2024shape}, then reconstruct and evaluate the masked region only.
In Figure~\ref{fig:nvs-dycheck}, both D3DGS~\cite{yang2024deformable} and L4GM often resolve a wrong scene scale or semi-transparent geometries, PGDVS~\cite{pgdvs} often creates broken geometry due to warping, while our method can handle them well.
Neither D3DGS nor PGDVS can correctly resolve the repetitive and fine-grained deformations in the paper windmill scene.
In addition, all optimization baseline methods are optimized with the entire sequence, while our DGS-LRM only uses a 24-frame continuous clip and 4 reference frames to reconstruct the scene.
The 4 reference frames are selected $\left[-96, -48, 48, 96\right]$ frames apart from the main input sequence, and we increase the number of frames in the opposite direction when meeting the start or end of the sequence.
In addition, the covisibility labeling of DyCheck is dedicated to methods optimizing with the full video sequence, which is disadvantageous to DGS-LRM with a small input window.
%

In Figure~\ref{fig:qual-davis}, we show that DGS-LRM can perform well even on in-the-wild videos. 
Compared with D3DGS~\cite{yang2024deformable}, we can correctly reconstruct the thin geometries of the bike wheels and challenging scenes with water deformation.
In flow visualizations, we illustrate that DGS-LRM effectively tracks the complex deformations in hand motion and wheel turning while maintaining consistent flow in the train's rigid body movements.
For visual clarity, we randomly downsample the per-pixel dense flows into sparse flows and then mask out the background flows using the object mask provided by DAVIS.

\vspace{\subsecmargin}
\subsection{3D Tracking}
\label{exp:tracking}
\vspace{\subsecmargin}

%
An accurate deformable 3D reconstruction should have a 3D deformation field aligned to the physically grounded 3D tracking trajectories.
In Table~\ref{tab:quant-track-pod} and Figure~\ref{fig:qual-pod}, we evaluate the quality of the reconstructed scene flow on the PointOdyssey benchmark~\cite{poointodyssey}. 
PointOdyssey includes 13 videos (ranging from 1{,}000 frames to 4{,}000 frames) of synthetic scenes with humanoid and animal meshes articulated with transferred real-world motions.
We compare DGS-LRM with the state-of-the-art SpatialTracker~\cite{xiao2024spatialtracker} and two baselines proposed in it.
DGS-LRM with flow chaining achieves comparable performance with SpatialTracker and performs better than other baselines.
In Figure~\ref{fig:qual-pod}, we found that SpatialTracker struggles on texture-less surfaces, where the tracking points drift over time in the regions with similar textures.
In contrast, as DGS-LRM prediction involves a total reconstruction of the object geometry with accurate depth prediction, it can better track points in completely texture-less areas.
For instance, the point registration on the ear root of the rabbit and the knee of the humanoid.
As discussed in Sec~\ref{method:chaining}, the chaining process involves failures when some points are completely invisible in a video segment. 
Therefore, we provide two additional variants: the native tracking performance without chaining (marked as Native) and an oracle case where we omit the flows with a significant discontinuity (threshold by L1 distance 0.1 meters) during the chaining process, which is marked as FC + FV.
While not directly comparable to SpatialTracker, these two variants show the actual quality of the predicted scene flows without the impairment caused by flow chaining.

\input{figures/pod}

\vspace{\subsecmargin}
\subsection{Ablation Study}
\label{exp:ablation}
\vspace{\subsecmargin}

\input{tables/ablation_combined}

In Table~\ref{tab:ablation}, we show that each proposed components contribute to the final performance.
The ablation is conducted at the 256$\times$256 resolution training stage, and we compare the rendering quality on the DyCheck benchmark.
The temporal tokenization enables training at scale by significantly reducing the memory consumption.
In Figure~\ref{fig:dual-view-compare}, we show that the dual-view sampling improves the training convergence on Kubric and leads to a better performance on our additionally rendered Kubric holdout test set (with new objects and HDRIs unseen during training).
The scene flow objective and reference frames both significantly boost the performance by a large margin.
In practice, we found the scene flow loss significantly improves the rigidity of the deformation, and the reference frames help solve the scene scale and depth by triangulating with distant views.

%% file: tables/quant.tex
\begin{table}[t]
\begin{minipage}{0.44\textwidth}
    \centering
    \scriptsize
    \renewcommand{\tabcolsep}{2pt}
    \captionof{table}{
        \textbf{Monocular Dynamic View Synthesis on DyCheck~\cite{dycheck}.}
        DGS-LRM outperforms LRM-based L4GM~\cite{ren2024l4gm}, and is comparable to optimization-based novel-view synthesis methods with a substantially faster reconstruction time.
        DynMask applies a dynamic mask to evaluate the foreground only.
    }
    \label{tab:quant-nvs-dycheck}
    \vspace{1.55em}
    \begin{tabular}{@{}lccccc@{}}
        \toprule
        Method &  Time (s) & DynMask & mPSNR ($\uparrow$) & mLPIPS ($\downarrow$) \\
        \midrule
        D3DGS~\cite{yang2024deformable} & 1-3 hours & $\times$ & 11.92 & 0.66 \\
        PGDVS~\cite{pgdvs}           & 3 hours & $\times$ & 15.88 & 0.34 \\
        Ours                    & 0.495 sec & $\times$ & 14.89 & 0.42 \\
        \midrule
        L4GM~\cite{ren2024l4gm} & 4.8 sec & $\checkmark$ & 5.84 & 0.67 \\
        Ours                    & 0.495 sec & $\checkmark$ & 11.97 & 0.51 \\
        \bottomrule
    \end{tabular}
    \vspace{-.5em}
    %
\end{minipage}
\hfill
\begin{minipage}{0.53\textwidth}
    %
    \renewcommand{\tabcolsep}{2pt}
    \scriptsize
    \captionof{table}{
        \textbf{3D Tracking on PointOdyssey~\cite{poointodyssey}.
    }
    DGS-LRM reconstructed 3D deformation field is comparable to state-of-the-art 3D tracking methods.
    %
    FC is flow chaining that combines scene flows from multiple segments. 
    FV is fully visible, which evaluates only the tracking points not occluded for more than 24 frames (DGS-LRM input length).
    }
    \label{tab:quant-track-pod}
    \vspace{.4em}
    \begin{tabular}{@{}lccccc@{}}
        \toprule
        Method & Frames & PSNR & ATE-3D ($\downarrow$) & $\delta_{0.1}$ ($\uparrow$) & $\delta_{0.2}$ ($\uparrow$) \\
        \midrule
        Chained RAFT3D~\cite{teed2021raft}   & 120 & N/A   & 0.70 & 0.12 & 0.25 \\
        Lifted CoTracker~\cite{karaev2024cotracker} & 120 & N/A   & 0.77 & 0.51 & 0.64 \\
        SpatialTracker~\cite{xiao2024spatialtracker}   & 120 & N/A   & 0.22 & \textbf{\underline{0.59}} & \textbf{\underline{0.76}} \\
        Ours (FC)                & 120 & 27.77 & \textbf{\underline{0.21}} & 0.57 & 0.68 \\
        \midrule
        Ours (Native)  &  24 & 27.77 & 0.11 & 0.72 & 0.84 \\
        Ours (FC + FV) & 120 & 27.77 & 0.15 & 0.64 & 0.75 \\
        \bottomrule
    \end{tabular}
    \vspace{-.5em}
\end{minipage}
\end{table}

%% file: figures/dycheck-nvs.tex
\begin{figure*}[t]
    \centering
    \setlength{\tabcolsep}{0pt}
    \begin{tabular}{cc ccccc c ccccc}
        %
        %
        \parbox[t]{1em}{\rotatebox[origin=l]{90}{\makecell{\scalebox{.7}{\hspace{1em} Time A}}}} &
        \includegraphics[width=.0805\linewidth]{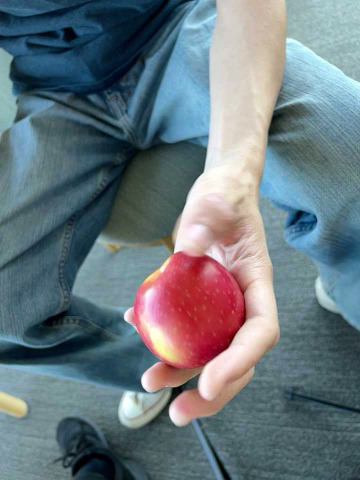} &
        \includegraphics[width=.0805\linewidth]{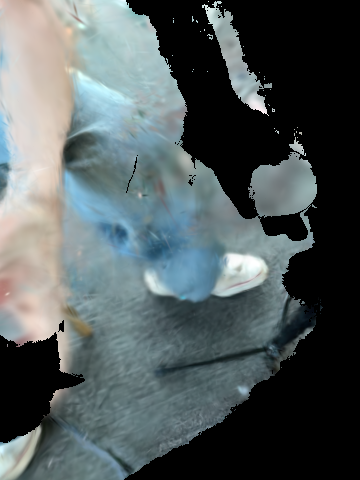} &
        \includegraphics[width=.0805\linewidth]{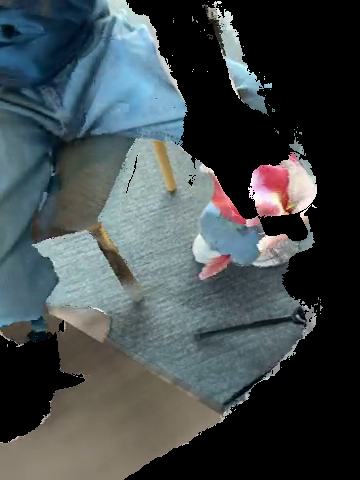} &
        \includegraphics[width=.0805\linewidth]{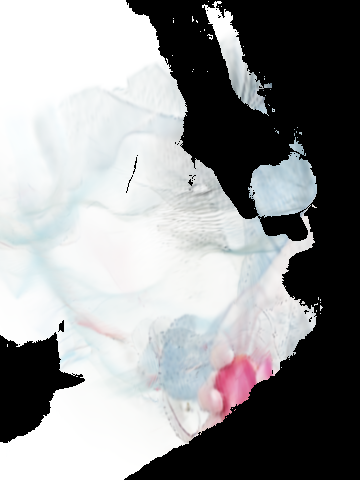} &
        \includegraphics[width=.0805\linewidth]{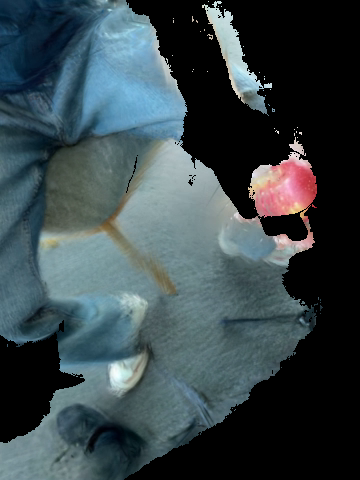} &
        \includegraphics[width=.0805\linewidth]{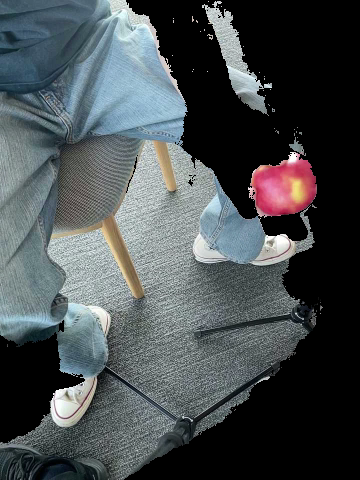} \hfill & \hfill
        \includegraphics[width=.0805\linewidth]{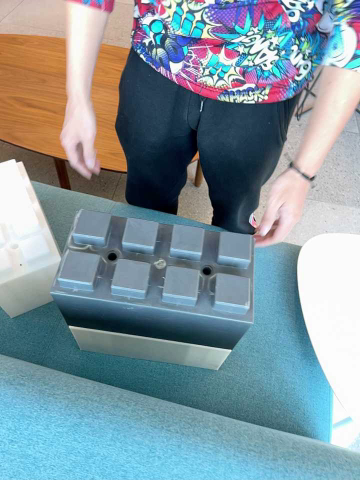} &
        \includegraphics[width=.0805\linewidth]{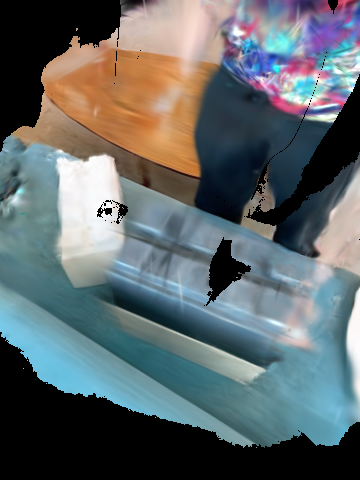} &
        \includegraphics[width=.0805\linewidth]{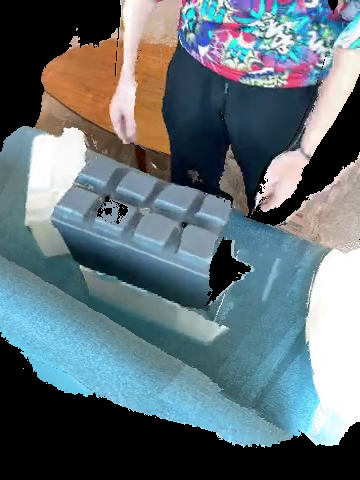} &
        \includegraphics[width=.0805\linewidth]{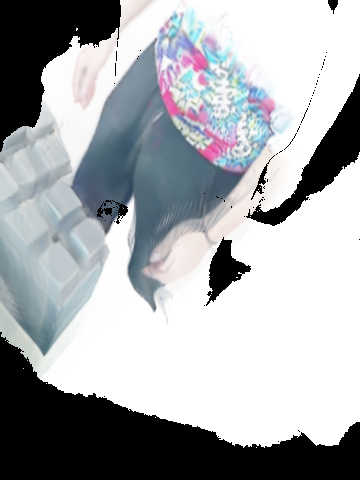} &
        \includegraphics[width=.0805\linewidth]{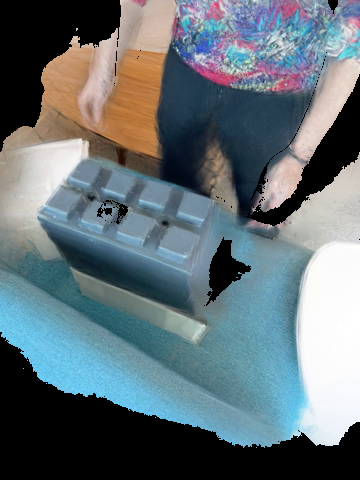} &
        \includegraphics[width=.0805\linewidth]{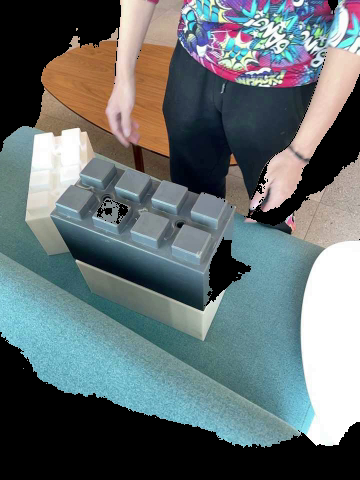}
        \\ [-0.25em]
        \parbox[t]{1em}{\rotatebox[origin=l]{90}{\makecell{\scalebox{.7}{\hspace{1em} Time B}}}} &
        \includegraphics[width=.0805\linewidth]{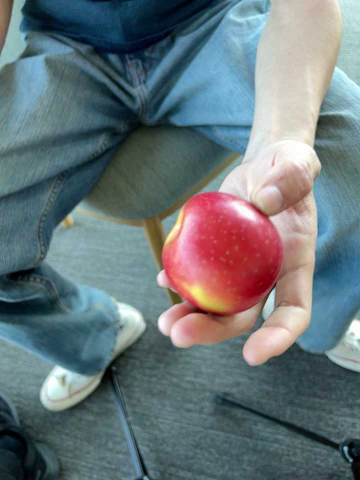} &
        \includegraphics[width=.0805\linewidth]{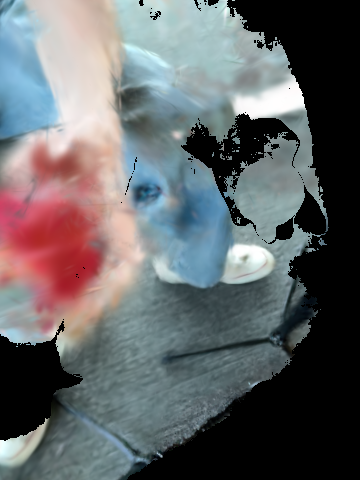} &
        \includegraphics[width=.0805\linewidth]{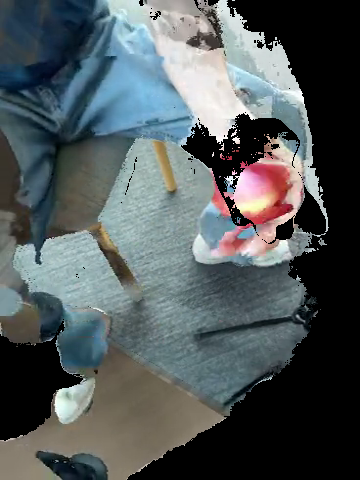} &
        \includegraphics[width=.0805\linewidth]{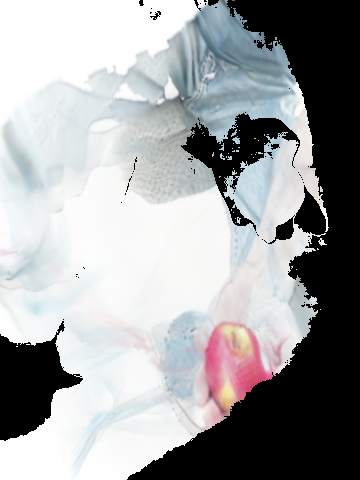} &
        \includegraphics[width=.0805\linewidth]{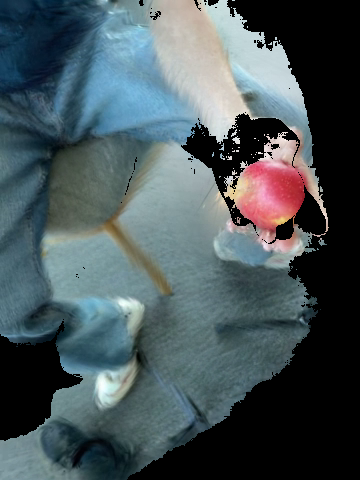} &
        \includegraphics[width=.0805\linewidth]{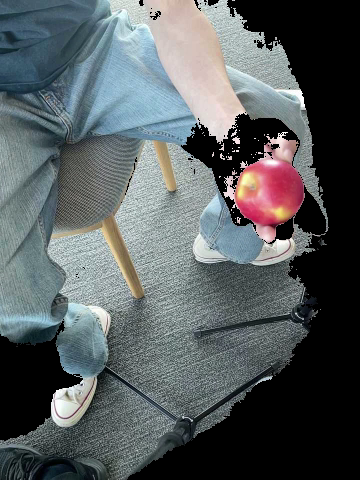} \hfill & \hfill
        \includegraphics[width=.0805\linewidth]{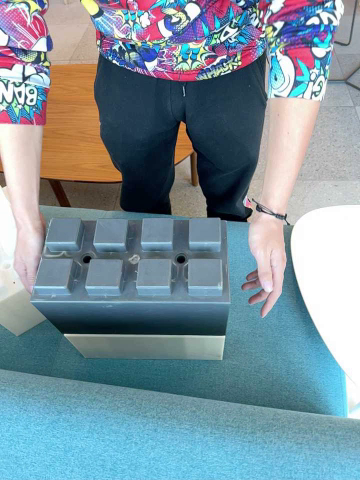} &
        \includegraphics[width=.0805\linewidth]{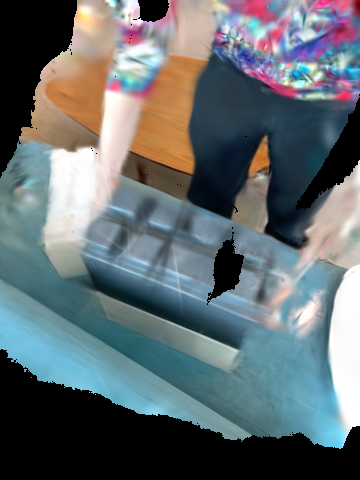} &
        \includegraphics[width=.0805\linewidth]{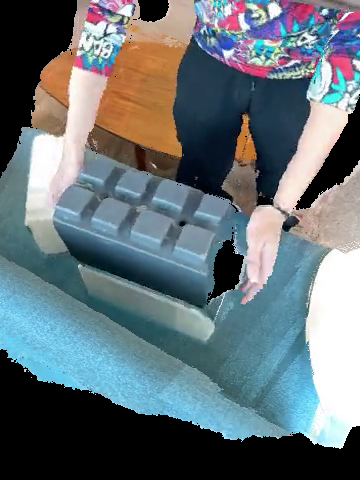} &
        \includegraphics[width=.0805\linewidth]{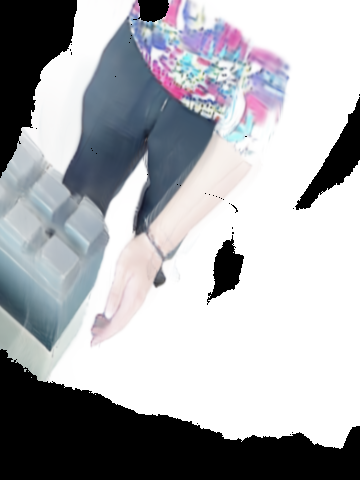} &
        \includegraphics[width=.0805\linewidth]{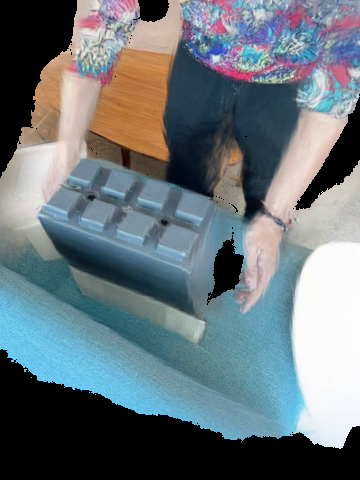} &
        \includegraphics[width=.0805\linewidth]{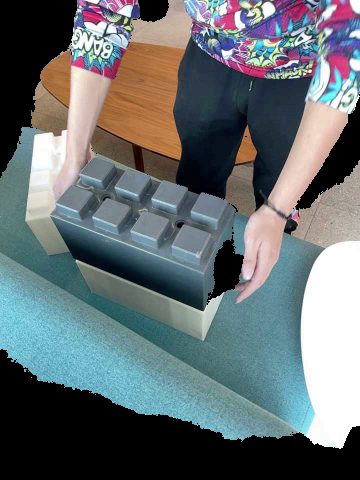} \\ [-0.25em]
        \midrule
        \parbox[t]{1em}{\rotatebox[origin=l]{90}{\makecell{\scalebox{.7}{\hspace{1em} Time A}}}} &
        \includegraphics[width=.0805\linewidth]{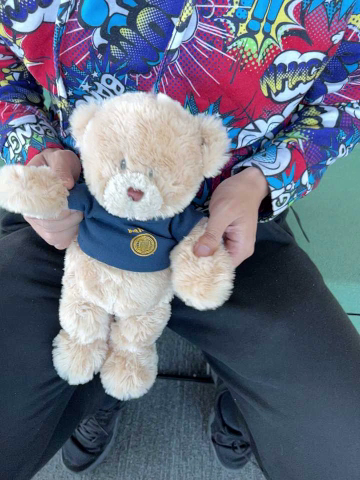} &
        \includegraphics[width=.0805\linewidth]{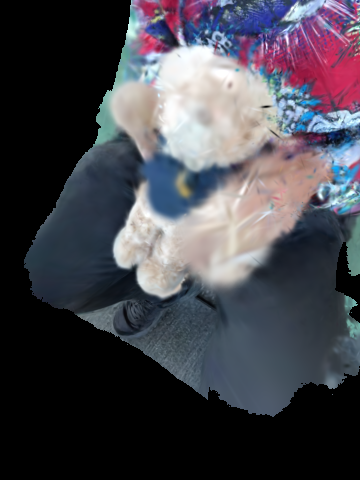} &
        \includegraphics[width=.0805\linewidth]{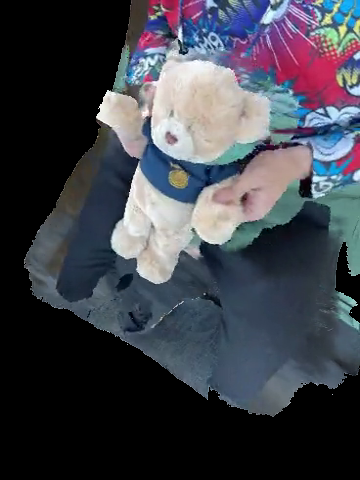} &
        \includegraphics[width=.0805\linewidth]{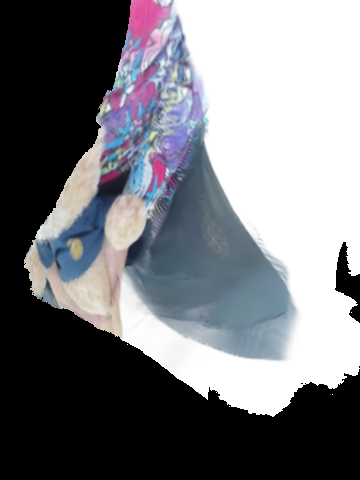} &
        \includegraphics[width=.0805\linewidth]{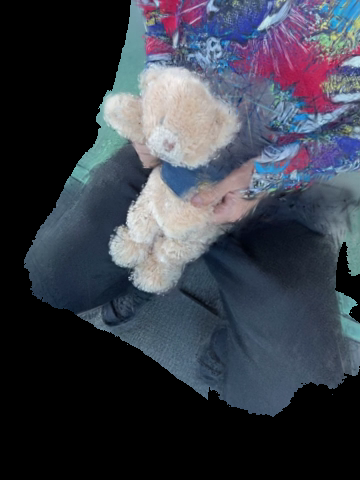} &
        \includegraphics[width=.0805\linewidth]{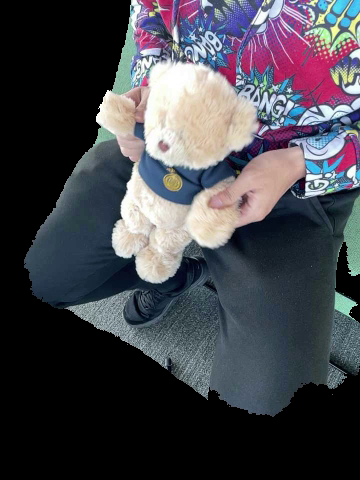} \hfill & \hfill
        \includegraphics[width=.0805\linewidth]{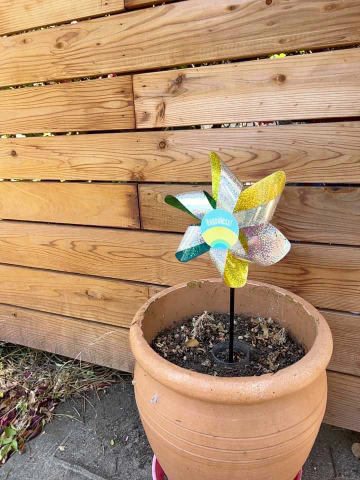} &
        \includegraphics[width=.0805\linewidth,trim={0cm 0cm 3cm 4cm},clip]{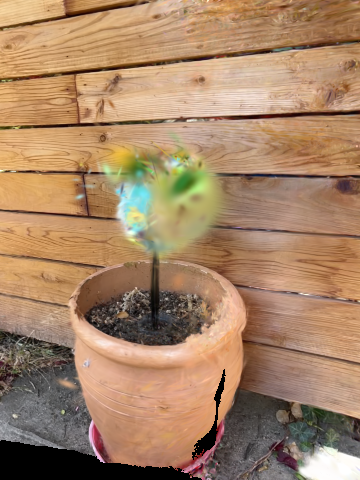} &
        \includegraphics[width=.0805\linewidth,trim={0cm 0cm 3cm 4cm},clip]{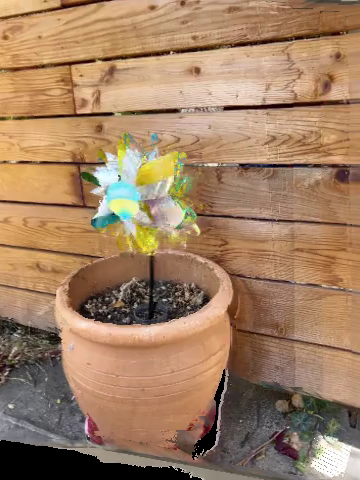} &
        \includegraphics[width=.0805\linewidth,trim={0cm 0cm 3cm 4cm},clip]{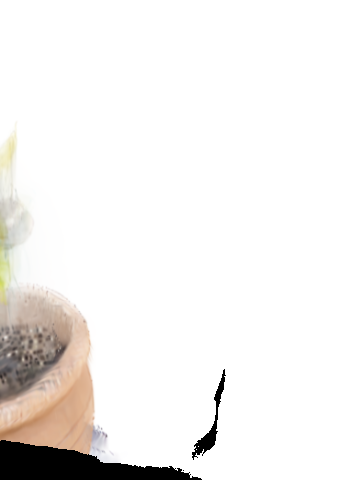} &
        \includegraphics[width=.0805\linewidth,trim={0cm 0cm 3cm 4cm},clip]{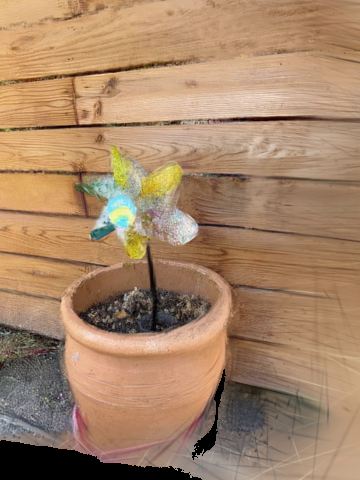} &
        \includegraphics[width=.0805\linewidth,trim={0cm 0cm 3cm 4cm},clip]{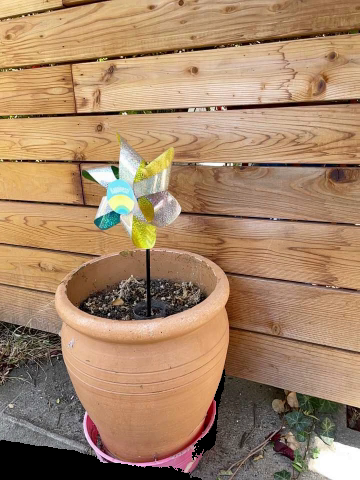} \\ [-0.25em]
        \parbox[t]{1em}{\rotatebox[origin=l]{90}{\makecell{\scalebox{.7}{\hspace{1em} Time B}}}} &
        \includegraphics[width=.0805\linewidth]{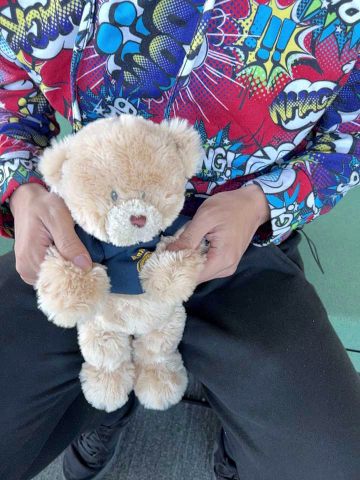} &
        \includegraphics[width=.0805\linewidth]{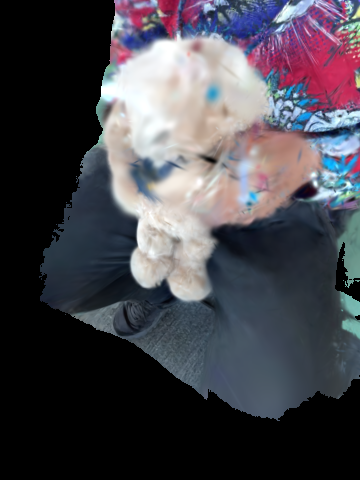} &
        \includegraphics[width=.0805\linewidth]{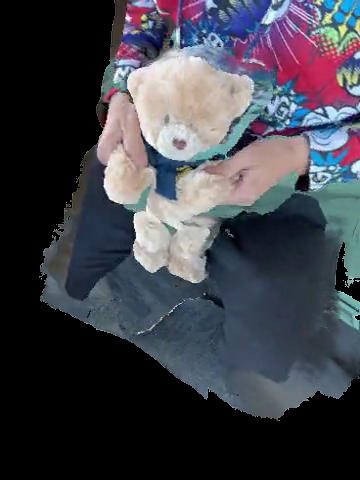} &
        \includegraphics[width=.0805\linewidth]{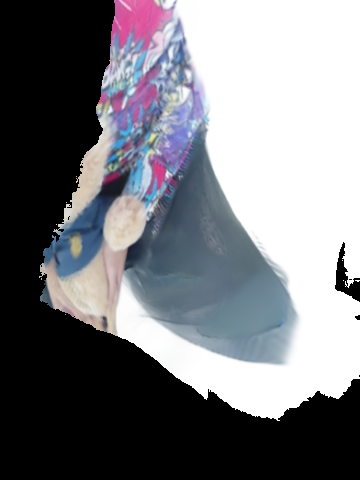} &
        \includegraphics[width=.0805\linewidth]{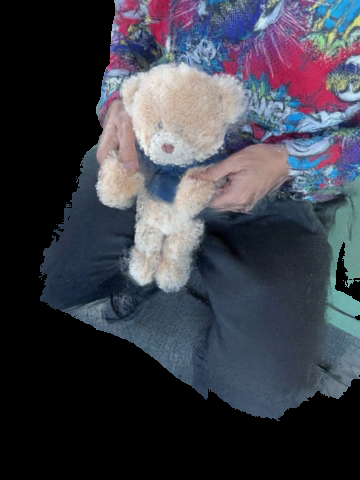} &
        \includegraphics[width=.0805\linewidth]{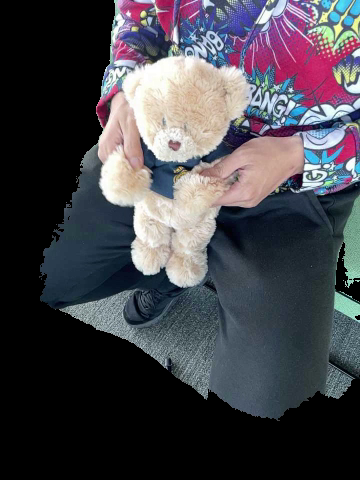} \hfill & \hfill
        \includegraphics[width=.0805\linewidth]{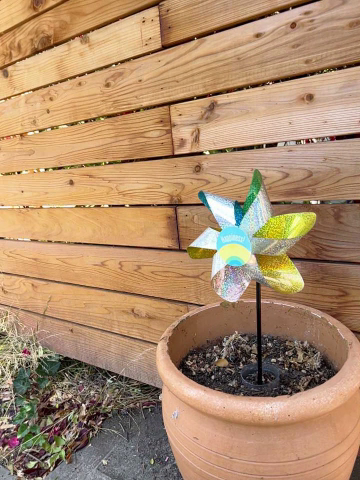} &
        \includegraphics[width=.0805\linewidth,trim={0cm 0cm 3cm 4cm},clip]{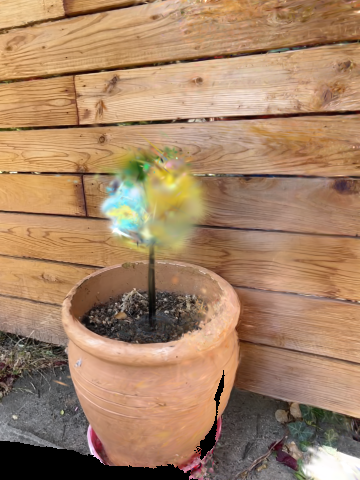} &
        \includegraphics[width=.0805\linewidth,trim={0cm 0cm 3cm 4cm},clip]{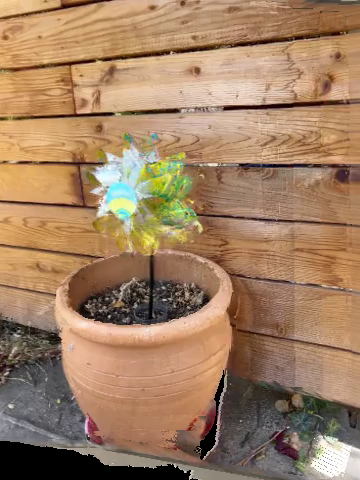} &
        \includegraphics[width=.0805\linewidth,trim={0cm 0cm 3cm 4cm},clip]{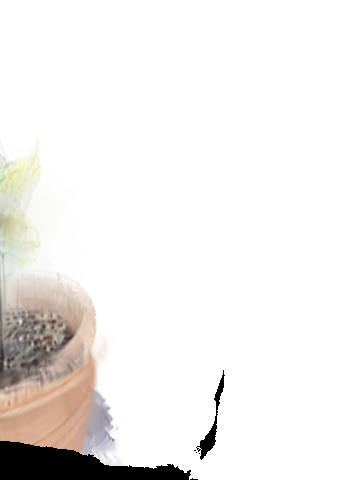} &
        \includegraphics[width=.0805\linewidth,trim={0cm 0cm 3cm 4cm},clip]{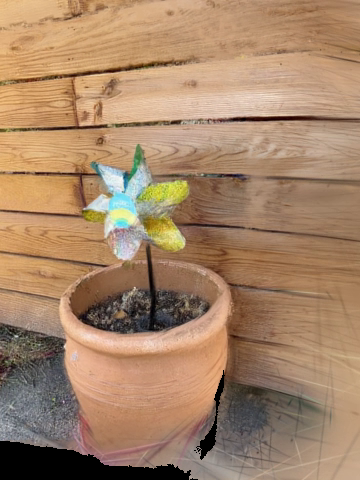} &
        \includegraphics[width=.0805\linewidth,trim={0cm 0cm 3cm 4cm},clip]{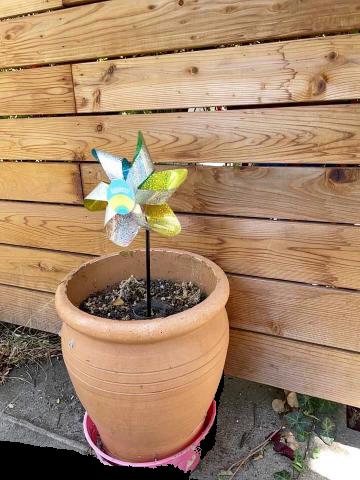} \\ [-0.35em]
        & \tiny Inputs & \tiny D3DGS & \tiny PGDVS & \tiny L4GM & \tiny Ours & \tiny GT & \tiny Inputs & \tiny D3DGS & \tiny PGDVS & \tiny L4GM & \tiny Ours & \tiny GT \\
    \end{tabular}
    \vspace{-.5em}
    \caption{
    \textbf{DyCheck iPhone dataset.} Our DGS-LRM outperforms D3DGS~\cite{yang2024deformable} and does not have warping artifacts as in PGDVS~\cite{pgdvs}. Both methods fail to recover the geometry and the repetitive motion of the windmill. We mask out the zero covisible regions with black pixels.
    }
    \label{fig:nvs-dycheck}
    \vspace{-.5em}
\end{figure*}

%% file: figures/davis.tex
\begin{figure*}[t]
    \centering
    \setlength{\tabcolsep}{0pt}
    \begin{tabular}{cc cc ccc}
        %
        %
        \includegraphics[width=.077\linewidth]{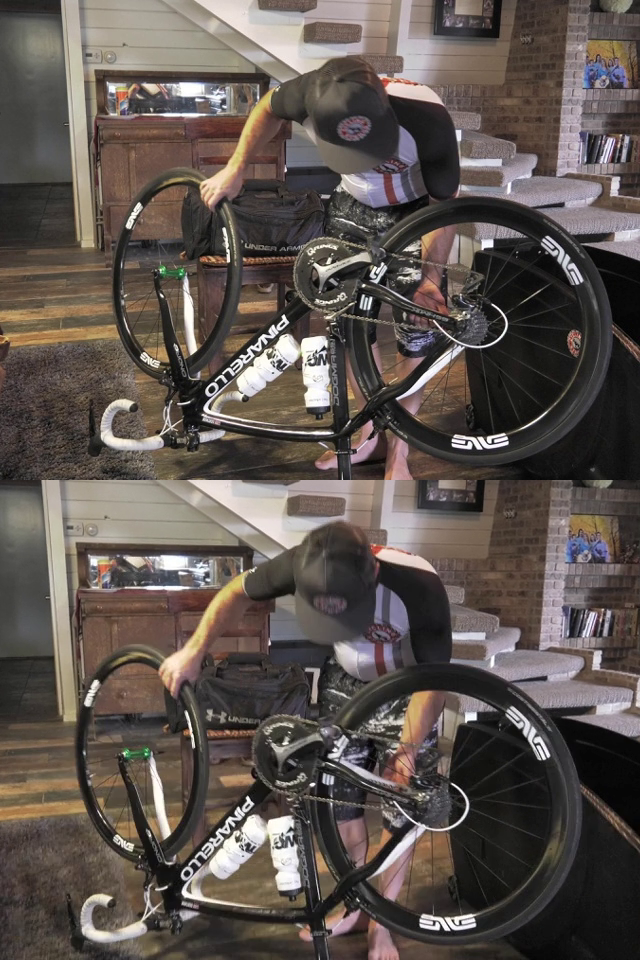} &
        \includegraphics[width=.154\linewidth]{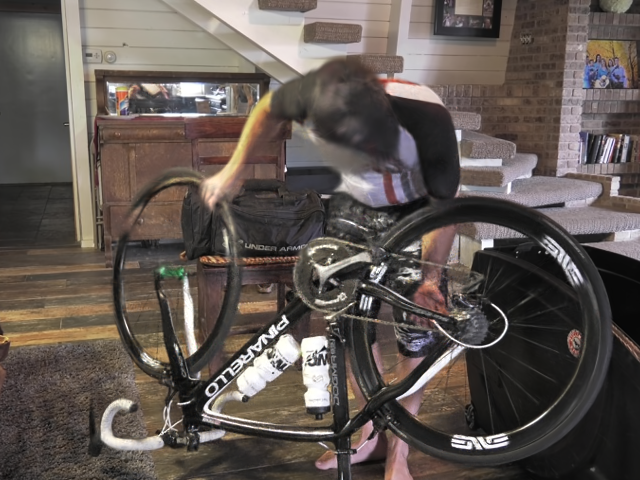} &
        \includegraphics[width=.154\linewidth]{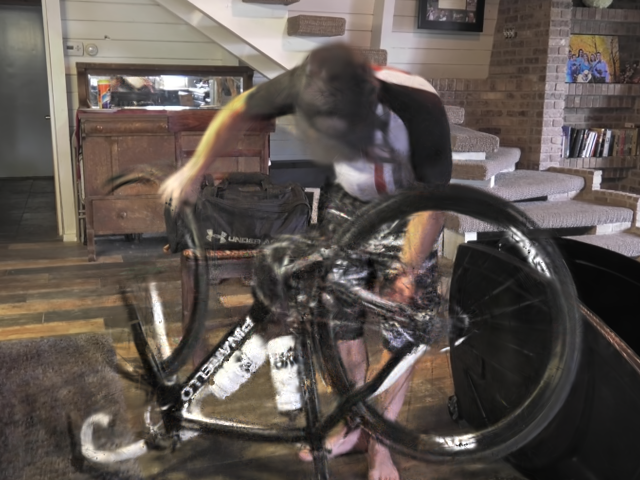} &
        \includegraphics[width=.154\linewidth]{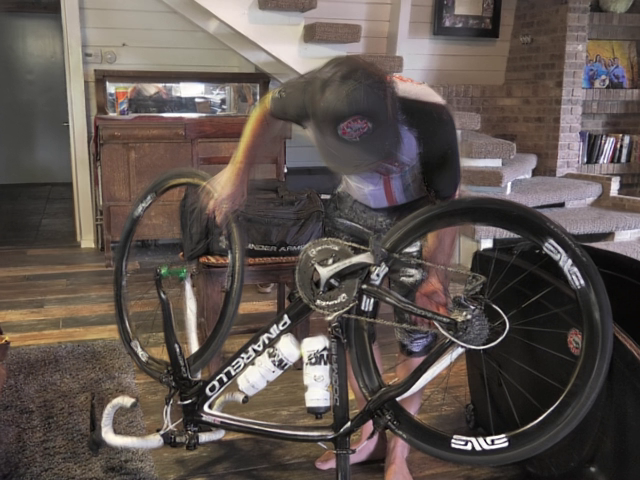} &
        \includegraphics[width=.154\linewidth]{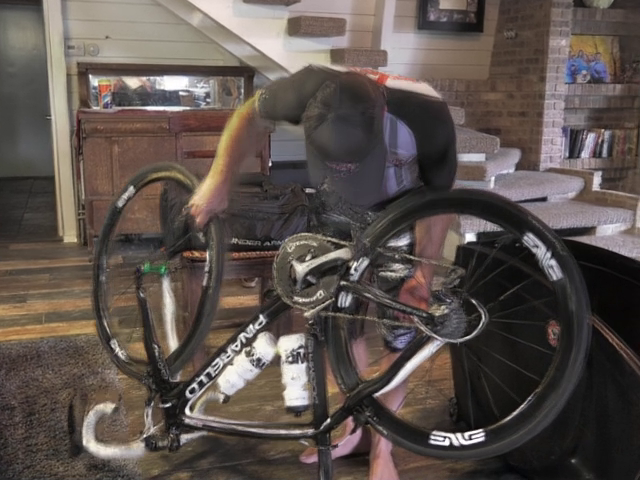} &
        \includegraphics[width=.154\linewidth]{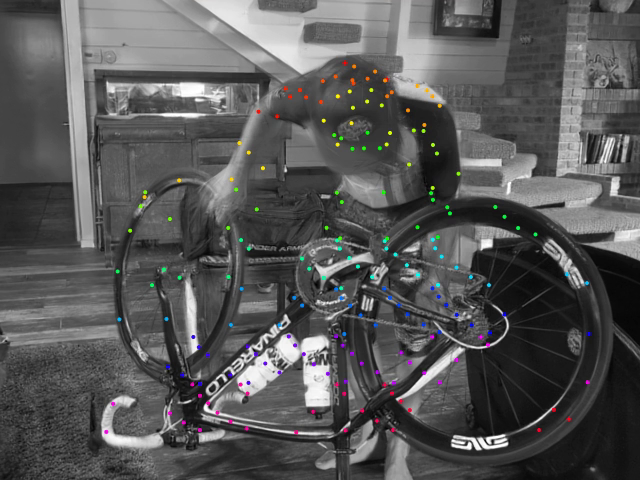} &
        \includegraphics[width=.154\linewidth]{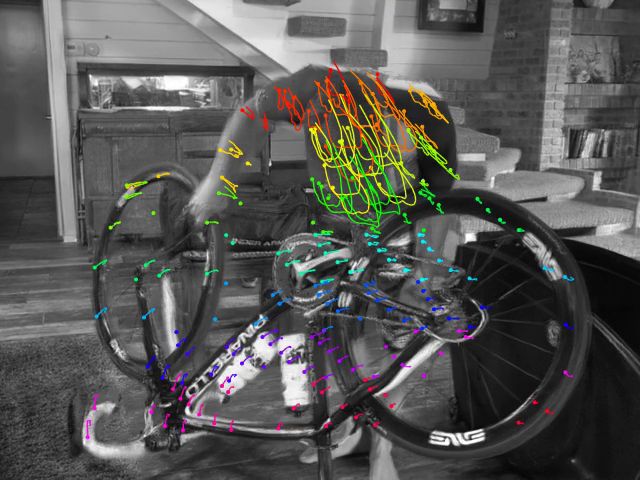}
        \\ [-0.25em]
        \includegraphics[width=.077\linewidth]{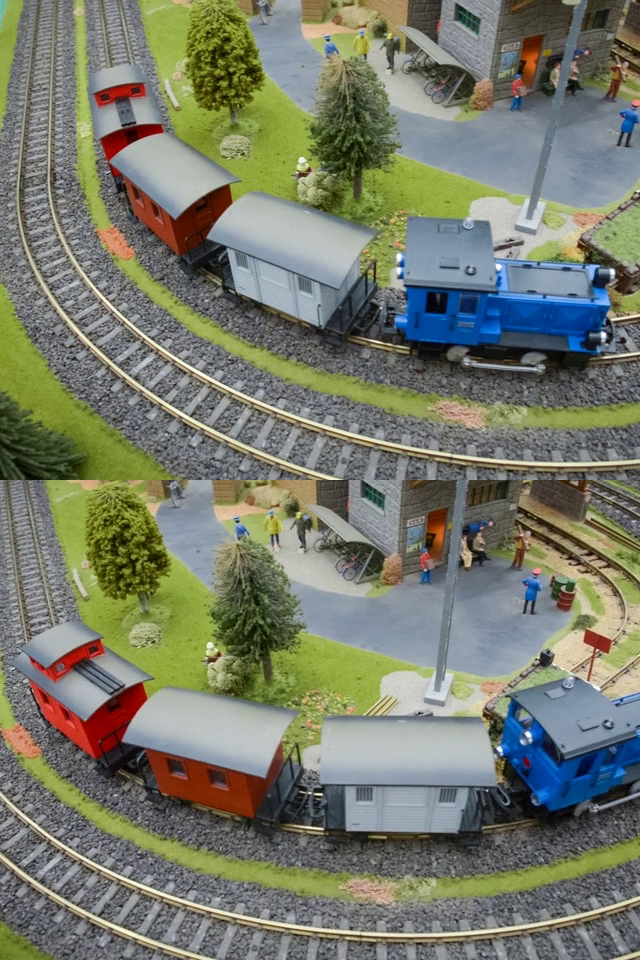} &
        \includegraphics[width=.154\linewidth]{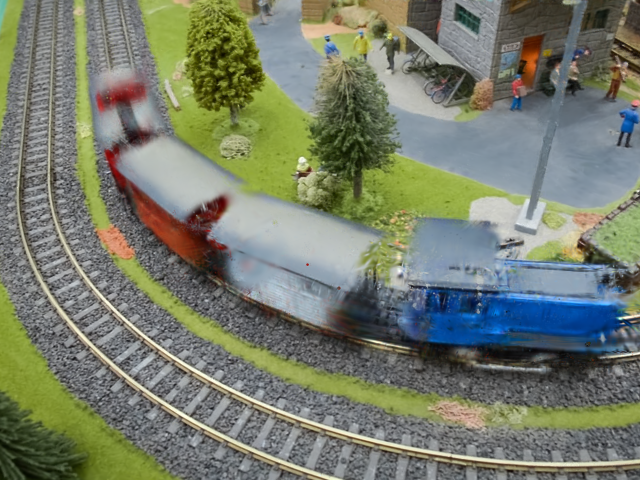} &
        \includegraphics[width=.154\linewidth]{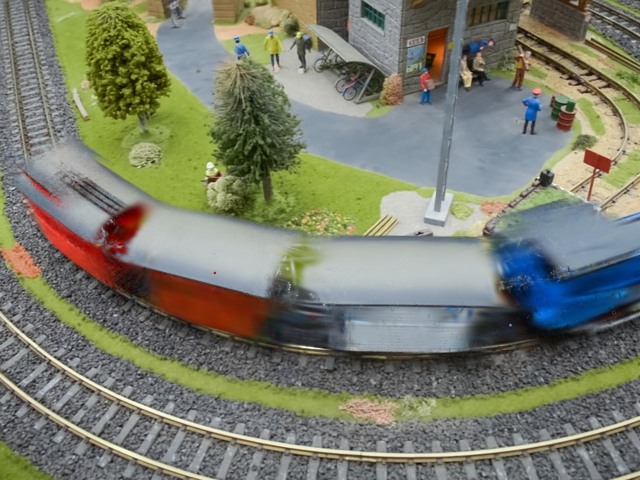} &
        \includegraphics[width=.154\linewidth]{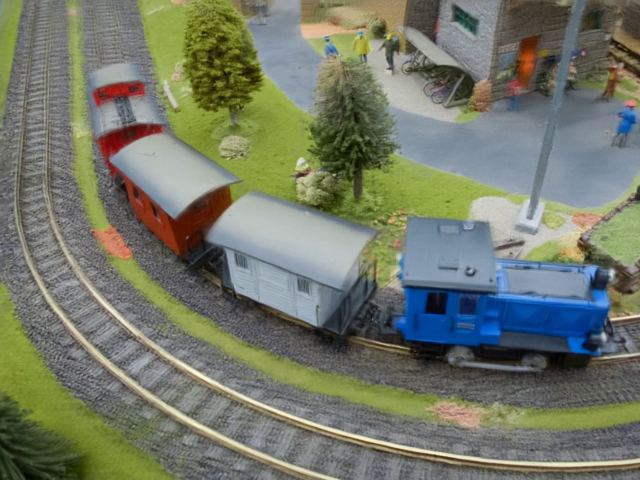} &
        \includegraphics[width=.154\linewidth]{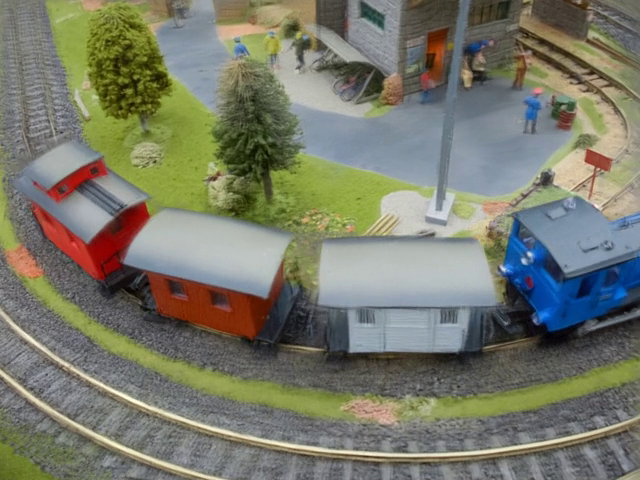} &
        \includegraphics[width=.154\linewidth]{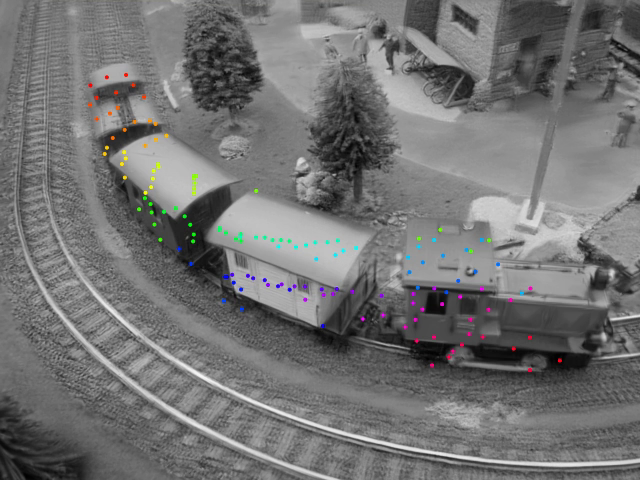} &
        \includegraphics[width=.154\linewidth]{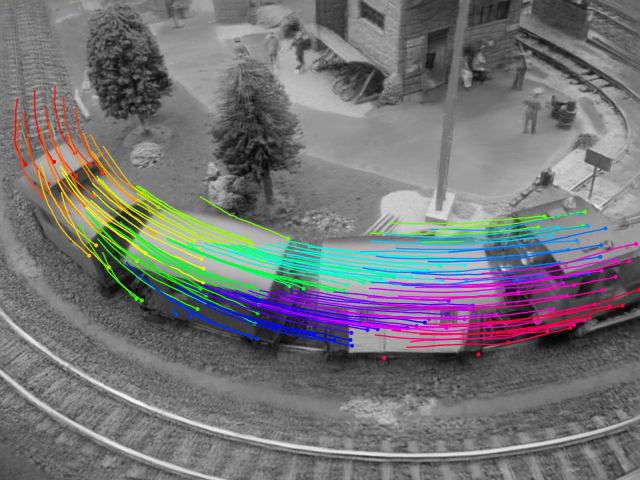}
        \\ [-0.25em]
        \includegraphics[width=.077\linewidth]{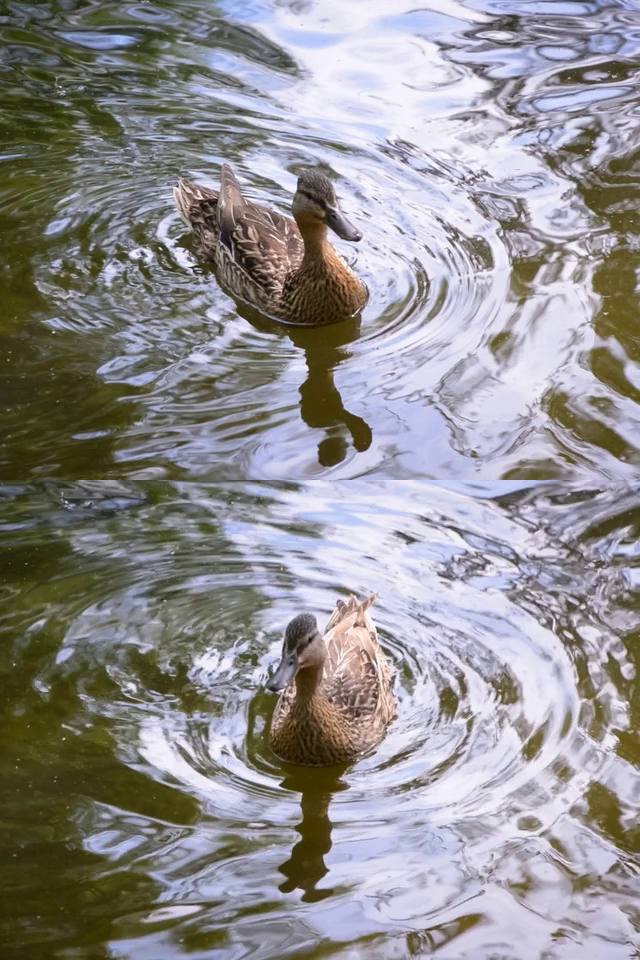} &
        \includegraphics[width=.154\linewidth]{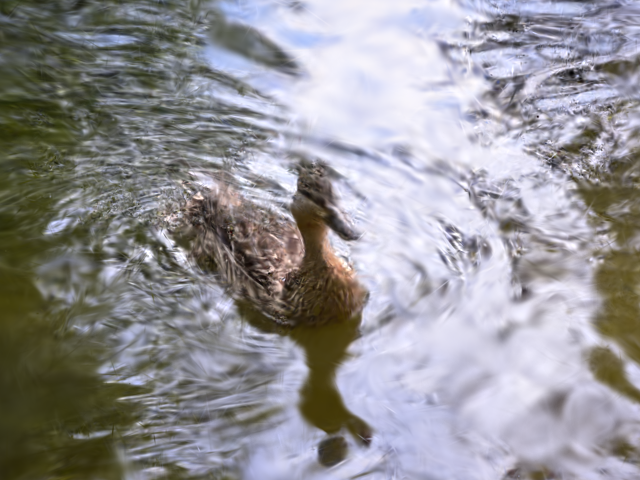} &
        \includegraphics[width=.154\linewidth]{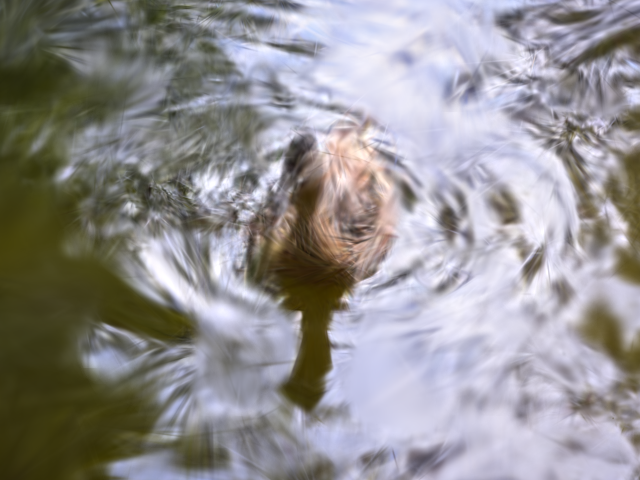} &
        \includegraphics[width=.154\linewidth]{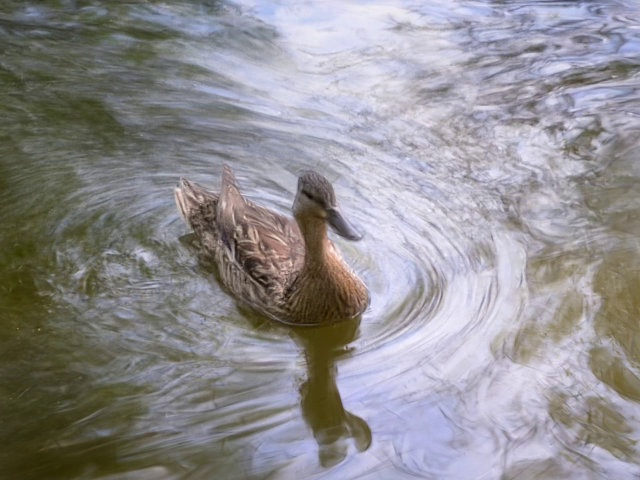} &
        \includegraphics[width=.154\linewidth]{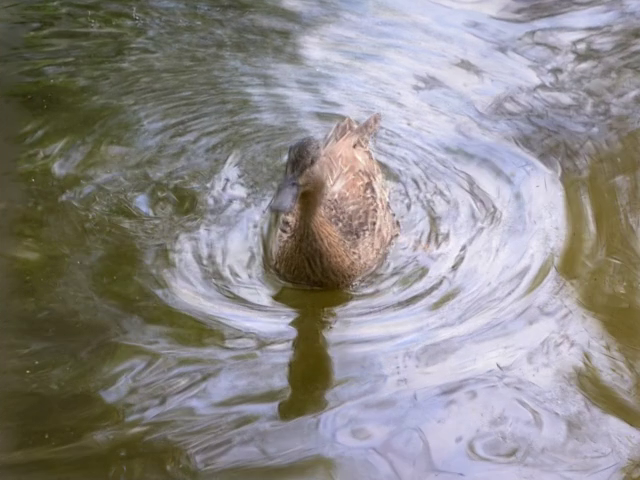} &
        \includegraphics[width=.154\linewidth]{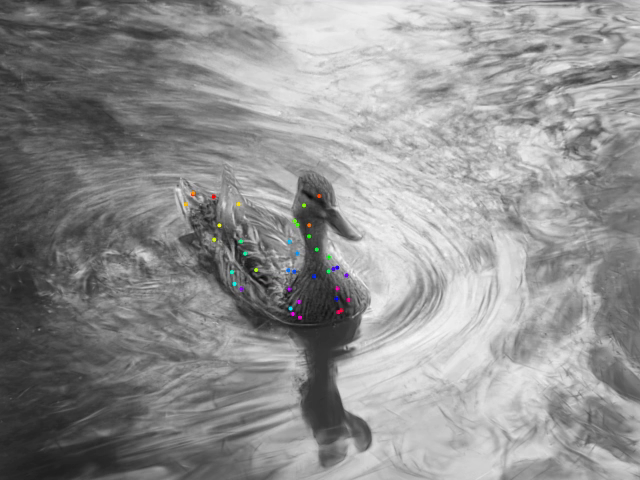} &
        \includegraphics[width=.154\linewidth]{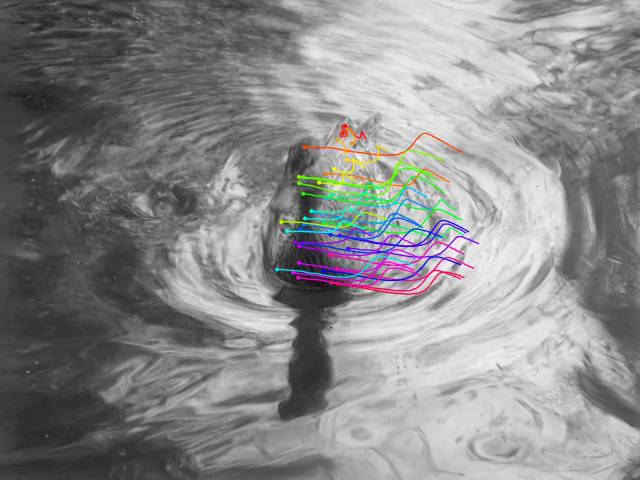}
        \\ [-0.25em]
        %
        \scriptsize Input Vid & \scriptsize D3DGS (Time A) & \scriptsize D3DGS (Time B) & \scriptsize Ours (Time A) & \scriptsize Ours (Time B) & \scriptsize Ours (Time A) & \scriptsize Ours (Time B) \\
    \end{tabular}
    \vspace{-.5em}
    \caption{
        \textbf{DAVIS dataset.} DGS-LRM outperforms D3DGS in synthesizing thin details and can predict consistent dynamic motion trajectories (visualized in color).
    }
    \label{fig:qual-davis}
    \vspace{-.5em}
\end{figure*}

%% file: figures/pod.tex
\begin{figure*}
    \centering
    \includegraphics[width=\linewidth]{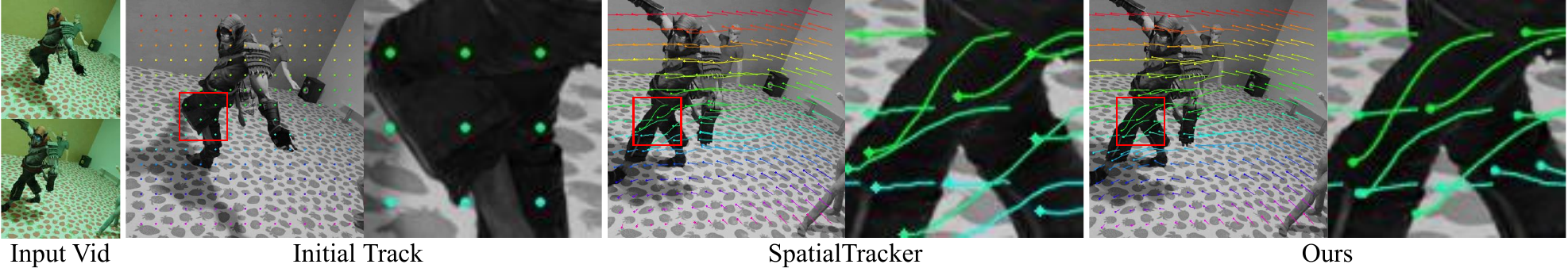} \vspace{-1.2em}
    \caption{
    \textbf{Qualitative comparisons to SpatialTracker\cite{xiao2024spatialtracker} on the PointOdyssey dataset.}
    DGS-LRM shows better performance and consistency in texture-less areas.
    SpatialTracker predicts certain tracks inconsistent with the object's moving direction. Such as several tracking points drift and collide in the humanoid's knee.
    }
    \vspace{-.7em}
    \label{fig:qual-pod}
\end{figure*}

%% file: tables/ablation_combined.tex
\begin{figure}[t]
\begin{minipage}{0.57\textwidth}
    \centering
    \scriptsize
    \renewcommand{\tabcolsep}{6pt}
    \captionof{table}{
    \textbf{Ablation Study.}
    We show that each component contributes to the final performance. 
    Note that the study is conducted with low-resolution models trained at 256$\times$256 resolution.
    (TT: temporal tokenization. DV: dual-view sampling. SF: scene flow loss. RF: reference frames.)}
    \begin{tabular}{@{}lccccc@{}}
        \toprule
        \multirow{2}{*}{Method} & \multicolumn{2}{c}{DyCheck} & \multicolumn{2}{c}{Kubric-MV (Test)} & \\
        \cmidrule(lr){2-3} \cmidrule(lr){4-5}
               & mPSNR ($\uparrow$) & mLPIPS ($\downarrow$) & mPSNR ($\uparrow$) & mLPIPS ($\downarrow$) \\
        \midrule
        w/o TT  & OOM   & OOM   & OOM   & OOM \\
        w/o DV  & \textbf{\underline{14.72}} & \textbf{\underline{0.412}} & 25.77 & 0.171 \\
        w/o SF  & 14.29 & 0.423 & 25.06 & 0.189 \\
        w/o RF  & 13.91 & 0.438 & 24.69 & 0.186 \\
        Full    & 14.67 & \textbf{\underline{0.412}} & \textbf{\underline{26.05}} & \textbf{\underline{0.161}} \\
        \bottomrule
    \end{tabular}
    \label{tab:ablation}
\end{minipage} %
\hfill%
\begin{minipage}{0.4\textwidth}
    \centering
    \scriptsize
    \centering
    \includegraphics[width=\textwidth]{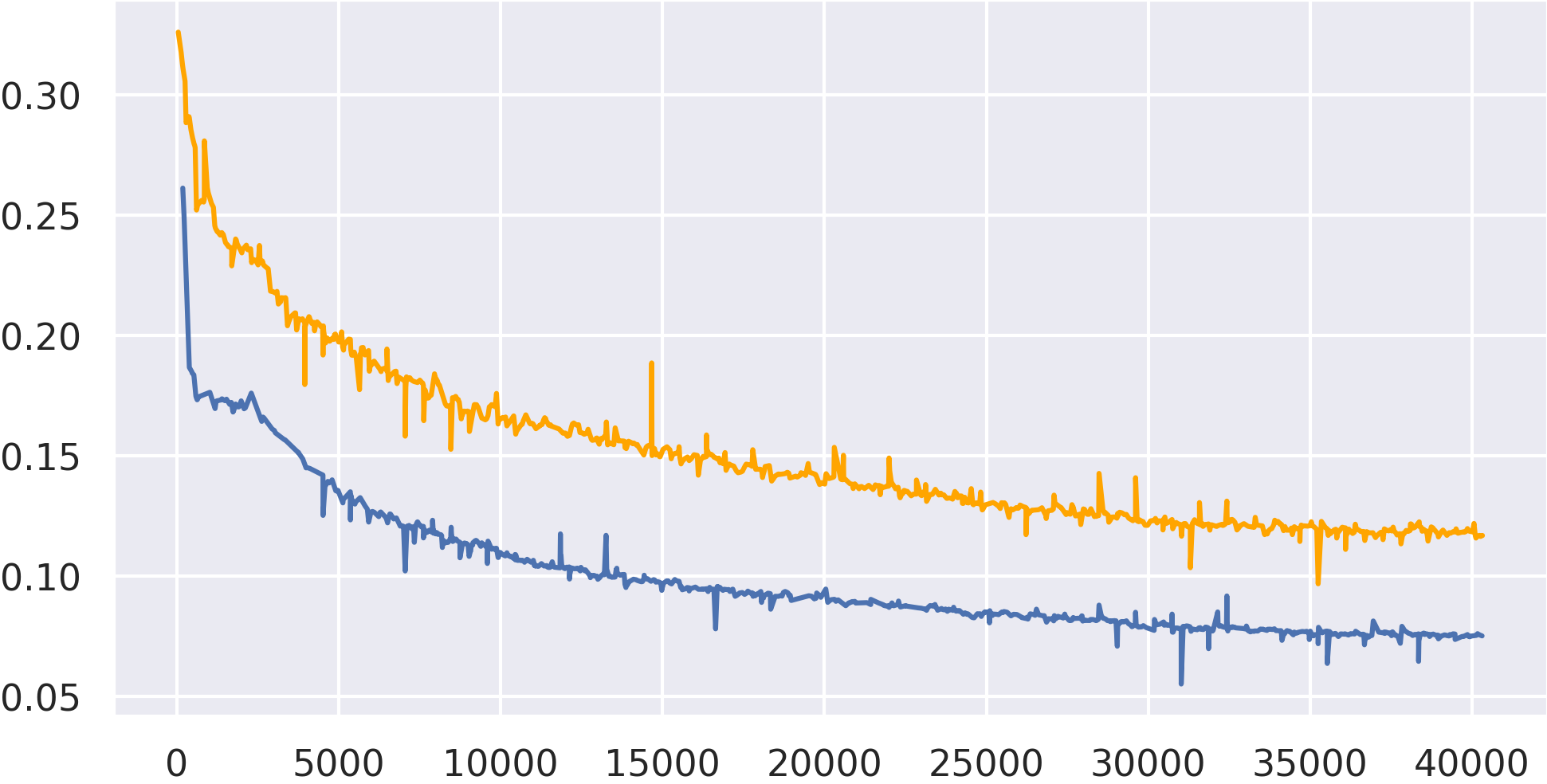}
    \vspace{-.5em}
    \captionof{figure}{
    Dual view supervision ({\color{blue} blue} curve) provides better training convergence in Perceptual Loss (LPIPS) compared to its counterpart ({\color{orange} orange} curve).
    }
    \label{fig:dual-view-compare}
\end{minipage}
\vspace{-1em}
\end{figure}

%% file: 6-conclusion.tex
\vspace{\secmargin}
\section{Conclusion and Limitations}
\vspace{\secmargin}

We introduce DGS-LRM, the first feed-forward network capable of predicting deformable 3D Gaussians from a posed monocular video in real-time. The predicted deformable Gaussians enable novel-view rendering, geometry reconstruction, and 3D scene flow estimation in world space. We train DGS-LRM on a large-scale multi-view rendered synthetic dataset and show that it generalizes well to real-world videos of varying complexity. Unlike prior monocular deformable 3D reconstruction methods, which require lengthy optimization to reconstruct and fuse priors from multiple individual networks, we demonstrate the potential of predicting deformable 3D Gaussians end-to-end and learning dynamic scene priors within a single network.

\Paragraph{Limitations.} DGS-LRM has a few limitations that can be explored in future works.
%
As the model is trained with temporally continuous video, the model cannot handle discrete image frames that are temporally too distant.
Our predicted scene flow cannot handle extremely large motion in the scene, which may stem from the motion distribution of the physically simulated synthetic dataset. Such a domain gap can also affect the synthesized novel view.
%
%
%
The input video baseline and distribution significantly influence the quality of novel view rendering quality. As the view deviates from the input trajectory, artifacts gradually intensify in the rendered images as well. \par